\documentstyle[12pt]{article}

\newcommand{\be}{\begin{equation}}
\newcommand{\ee}{\end{equation}}
\newcommand{\ba}{\begin{eqnarray}}
\newcommand{\ea}{\end{eqnarray}}
\newcommand{\x}{\vec{x}}
\newcommand{\k}{\vec{k}}
\newcommand{\vP}{\vec{\Phi}}
\newcommand{\vp}{\vec{\pi}}

\newcommand{\h}{\hbar}
\begin{document}
\begin{center}
{\bf  NON-EQUILIBRIUM DYNAMICS OF PHASE TRANSITIONS:}
{\bf FROM THE EARLY UNIVERSE TO CHIRAL CONDENSATES}
\end{center}
\begin{center}
{\bf D. Boyanovsky$^{(a)}$, H.J. de Vega$^{(*b)}$ , R. Holman$^{(c)}$,}\\
\end{center}
\begin{center}
{\it (a)  Department of Physics and Astronomy, University of
Pittsburgh, Pittsburgh, PA. 15260, U.S.A.} \\
{\it (b)  Laboratoire de Physique Th\'eorique et Hautes Energies$^{[*]}$
Universit\'e Pierre et Marie Curie (Paris VI),
Tour 16, 1er. \'etage, 4, Place Jussieu
75252 Paris, Cedex 05, France}\\
{\it (c) Department of Physics, Carnegie Mellon University, Pittsburgh,
PA. 15213, U. S. A.}
\end{center}
\begin{abstract}
In this brief review we introduce the methods of quantum field theory out
of equilibrium and study the non-equilibrium aspects of phase transitions.
Specifically we critically study the picture of the ``slow-roll'' phase
transition in the new inflationary models, we show that the instabilities
that are the hallmark of the phase transition, that is the formation of
correlated domains, dramatically change this picture. We analyze in
detail the dynamics of phase separation in strongly supercooled phase
transitions in Minkowski space. We argue that this is typically the
situation in weakly coupled
scalar theories. The effective evolution equations for the expectation
value and the fluctuations of an inflaton field
in a FRW cosmology are derived both in the loop expansion and in a
self-consistent non-perturbative scheme.
Finally we use these non-equilibrium techniques
and concepts to study the influence of quantum and thermal fluctuations on
the dynamics of a proposed mechanism for the formation of disoriented
chiral condensates during a rapid phase transition out of equilibrium. This
last topic may prove to be experimentally relevant at present accelerator
energies.

{\it To appear in the Proceedings of the `2nd. Journ\'ee Cosmologie',
Observatoire de Paris, 2-4, June 1994. H J de Vega and N. S\'anchez,
Editors, World Scientific.}
\end{abstract}

\newpage

\section{Introduction and Motivation}

Phase transitions play a fundamental role in our understanding of
the interplay between cosmology and particle physics in extreme
environments. It is widely
accepted that the Universe that we see today is a remnant of
 several different phase transitions that took place
at different energy (temperature) scales and with remarkable
consequences at low temperatures and energies.
Restricting ourselves to the domain of particle physics, there
is the conjecture that a Grand Unified phase transition has taken
place at about $10^{15}$ Gev, in which a GUT symmetry breaks
down to $SU(3)\otimes SU(2)\otimes U(1)$. Within the desert hypothesis, the
next phase transition took place at about 240 Gev at which the remaining
symmetry is broken down  to $SU(3)\otimes U(1)|_{em}$. Finally at
a scale of about 100-200 Mev the confining and chiral phase transition
ocurred. Heavy Ion colliders at RHIC and perhaps LHC will probe this
energy region within the next few years, thus the understanding of
the non-equilibrium properties of phase transitions becomes a more
tangible problem.

To put things in perspective, the temperature at the
center of the sun is about $10^7 K \approx 1 keV$.

 There is
also the possibility that the observed baryon asymmetry
 may be explained as a result of a phase transition at electroweak energy
scales and non-equilibrium conditions \cite{andrei}.

Phase transitions are an essential ingredient in inflationary models of
the early universe\cite{guth1}-\cite{abbott}.
The importance of the description of phase transitions in
extreme environments was recognized long time ago and efforts were
devoted to their description in relativistic
quantum field theory at finite temperature\cite{kirzhnits}-\cite{sweinberg}.
 For a very thorough account of  phase transitions in the early universe
 see the reviews by Brandenberger\cite{brandenberger},
 Kolb and Turner\cite{kolb} and Linde\cite{linde}.

 The methods used to study the {\it equilibrium} properties of
 phase transitions, like the order of the transition and critical temperature
 are by now well understood and widely used,
 in particular field theory at finite temperature and effective
 potentials\cite{kapusta}.

 These methods, however, are restricted to a {\it static} description
 of the consequences of the phase transition, but can hardly be used to
 understand the {\it dynamics} of the processes involved  during
 the phase transition. In particular, the effective potential
 that is widely used to determine the nature of a phase transition and
 static quantities like critical temperatures etc, is {\bf not} the
appropriate tool
 for the description of the dynamics. The effective potential corresponds
 to the {\it equilibrium} free energy density as a function of the
 order parameter. This is a static quantity, calculated in equilibrium, and
  to one loop order it is complex within the region of
 homogeneous field
 configurations in which $V''(\phi)< 0$, where $V(\phi)$ is the classical
 potential. This was already recognized in the early treatments by Dolan
 and Jackiw\cite{dolan}.

 In statistical mechanics, this region is referred to as the
 ``spinodal'' and corresponds to a sequence of states which are
  thermodynamically unstable.

  At zero temperature, the imaginary part of
  the effective potential has been identified with the decay rate of this
  particular unstable state\cite{weinbergwu}.

The use of the static effective potential to describe the dynamics of
phase transitions has been criticized by many authors \cite{mazenko,boyveg}.
 These authors argued that phase transitions in
typical theories will occur  via the formation and growth of
correlated domains inside which the field will relax to the value of the
minimum of the equilibrium free energy fairly quickly. It is now believed
that this picture may be correct for {\it strongly coupled theories} but
is not accurate for weak couplings.

The mechanism that is responsible for a typical second order
phase transition from an initially symmetric high temperature phase is
fairly well known. When the temperature becomes lower than the critical
temperature, long-wavelength fluctuations become unstable and
begin to grow and the field becomes correlated inside ``domains''.  The order
parameter (the expectation value of the volume average of the field),
remains zero all throughout the transition. These correlated regions grow
as a function of time and eventually at long times (determined by
non-linear and
dissipative
processes) the system reaches a state of equilibrium. Inside each correlation
volume the field is at a minima of the effective potential.

This is the process of phase separation, whether this process can be studied
in equilibrium or not is a matter of time scales.

{\bf Equilibrium or Non-Equilibrium: C'est la question!}

Consider the situation of an homogeneous and isotropic cosmological
model in which the temperature of the universe decreases because
of the expansion typically as
\begin{equation}
T(t)= T(t_0)\,{{a(t_0)}\over{a(t)}}
\end{equation}
with $a(t)$ the cosmological expansion  factor.  The cooling rate is
thus determined by
\begin{equation}
-\dot{T}/T = H  \label{coolingrate}
\end{equation}
where $H$ stands for the Hubble constant, $ H(t)\equiv \dot{a(t)}/a(t)$.
This determines the time scale for variation of temperature. If relaxational
processes of the system are much faster than this scale then the system is
in ``quasi-thermodynamic equilibrium'' locally. Then a determination of whether
an equilibrium description is valid requires the comparison of these two time
scales.

The typical collisional
relaxation rate for a process at energy E in the heat bath
is given by
$\Gamma(E) = \tau^{-1}(E) \approx n(E)\sigma(E)v(E)$ with
$n(E)$ the number
density of particles  with this energy E, $\sigma(E)$ the
scattering cross
section at this energy and $v(E)$ the velocity of the
incident beam of
particles. The lowest order (Born approximation) scattering
cross
section in a typical
scalar theory with quartic interaction is $\sigma(E) \approx
\lambda^2 / E^2$.
At very high temperatures, $T \gg m_{\Phi}$, with $m_{\Phi}$
the mass of the
field, $n(T) \approx T^3$, the internal energy density is
$U/V \approx T^4$,
and the average energy per particle is $\langle E \rangle
\approx T$, and
$v(T) \approx 1$.
Thus the typical collisional relaxation rate is
$\Gamma(T) \approx \lambda^2 T$. In an expanding universe,
the conditions for
local equilibrium will prevail provided that $\Gamma(T) \gg H(t)$,
with $a(t)$ the FRW scale factor. If this is the case, the
collisions occur  very quickly compared to the expansion rate, and
particles will
equilibrate. This argument applies to the collisional relaxation of high
frequency (short-wavelength) modes for which $k \gg m_{\Phi}$. This
may be understood as follows. Each ``external leg'' in the scattering process
considered, carries  typical momentum and energy $k\approx E \approx T > T_c $.
But in these typical theories $T_c \approx m_{\Phi}/
\sqrt{\lambda}$.  Thus for weakly coupled theories $T_c \gg m_{\Phi}$.

To obtain an order of magnitude estimate, we concentrate
near the phase transition at $T \approx T_c \approx 10^{14}
{\mbox{Gev}}$, $H \approx T^2 / M_{Pl} \approx 10^{-5}\; T$,
this implies that for $\lambda > 0.01$ the conditions for
local equilibrium may prevail. However, in weakly coupled
inflaton models of
inflation, phenomenologically the coupling is bound by the
spectrum of density fluctuations to be $\lambda \approx
10^{-12}-10^{-14}$\cite{linde,starobinsky}. Thus, in these weakly
coupled theories  the
conditions for local equilibrium of high energy modes may
not be achieved. One may, however, assume that although the
scalar field
is weakly self-coupled, it has strong coupling to the heat
bath (presumably
other fields in the theory) and thus remains in local
thermodynamic equilibrium.

This argument, however, applies to the collisional
relaxation of {\it short
wavelength modes}. We observe, however, that these type of
arguments are
{\it not valid} for the dynamics of the {\it long-wavelength
modes} at
temperatures {\it below} $T_c$, for the following reason.

At very high temperatures, and in local equilibrium, the
system is in the
disordered phase with $\langle\Phi\rangle = 0$ and {\it
short ranged
correlations}, as measured
by the equal time correlation function (properly subtracted)
\begin{eqnarray}
\langle\Phi(\vec{r},t)\Phi(\vec{0},t)\rangle & \approx & T^2
\exp[-
|\vec{r}|/ \xi(T)]
\label{correlationfunction} \\
\xi(T)                           & \approx &
\frac{1}{\sqrt{\lambda} T}
\approx \xi(0)\;\frac{T_c}{T} \label{hitcorrelationlength}
\end{eqnarray}

As the temperature drops near the critical temperature, and
below, the phase
transition occurs. The onset of the phase transition is
characterized by the
instabilities of long-wavelength fluctuations, and the
ensuing growth of
correlations. The field begins to correlate over larger
distances, and ``domains'' will form and grow.
 If the
initial value of the order parameter is zero, it will remain
zero throughout the transition.
This process is very similar to the
process of spinodal decomposition, or phase separation in condensed matter
systems\cite{langer,guntonmiguel}. This
growth of correlations, cannot be described as a process in
local thermodynamic equilibrium.

These instabilities are
manifest in the {\it equilibrium} free energy in the form of
imaginary parts, and the equilibrium free energy is not a
relevant quantity to study the {\it dynamics}.

These long
wavelength modes whose instabilities trigger the phase
transitions have very
slow dynamics. This is the phenomenon of critical slowing down that is
observed experimentally in binary mixtures and numerically in typical
simulations of phase transitions\cite{stanley,langer}. The long wavelength
fluctuations correspond to coherent collective behavior in
which degrees of freedom become correlated over large
distances. These collective long-wavelength modes have
extremely
slow relaxation near the phase transition, and they do
not have many available low energy decay channels.
Certainly through the phase transition, high frequency,
short wavelength modes
may still remain in local equilibrium by the arguments
presented above (if the coupling is sufficiently strong),
they
have many channels for decay and thus will maintain local
equilibrium through the phase transition.

To make this argument more quantitative, consider the
situation in which the final temperature is below the
critical value and early times after the transition. For
small amplitude fluctuations of the field, long wavelength
modes ``see'' an inverted harmonic oscillator and the
amplitude fluctuations begin to grow as (see below)
\begin{eqnarray}
\langle\Phi_{k}(t)\Phi_{-\vec{k}}(t)\rangle & \approx &\exp[2
W(k)t]
\label{unstable} \\
W(k)                      &  =      & \sqrt{\mu^2(T)-
\vec{k}^2}
\label{bigomega} \\
\mu^2(T)                  &   =     & \mu^2(0)\;\left[1-
\left(\frac{T_f}{T_c}\right)^2\right]
\end{eqnarray}
for $\vec{k}^2 < \mu^2(T)$.

In particular this situation, modelled with the inverted harmonic
oscillators is precisely the situation thoroughly and clearly studied by
 Weinberg and Wu\cite{weinbergwu} and Guth and Pi\cite{guthpi} .

The time scales that must be compared for the dynamics of these instabilities,
are now the growth rate $ \Gamma(k) \approx \sqrt{\mu^2(T)-\vec{k}^2}$ and
the expansion rate $H \approx T^2/M_{Pl} \approx \; 10^{-5} \; T $ , if the
expansion rate is comparable to the growth rate, then the long wavelength
modes that are unstable and begin to correlate may be in local thermodynamic
equilibrium through the cooling down process. Using $ T_c \approx \mu(0) /
\sqrt{\lambda}$, we must compare $ [1-(\frac{T}{T_c})^2]^{\frac{1}{2}}$ to
$10^{-5}/ \sqrt{\lambda}$. Clearly for weakly coupled theories, or
very near the critical temperature, the growth rate of the
unstable modes will be much slower than the rate of cooling down, and the
phase transition will be supercooled, similarly to a ``quenching process''
from a high temperature, disordered phase to a supercooled
low temperature  situation. For example, for $\lambda \approx 10^{-12}$  the
growth rate of long wavelength fluctuations, is much smaller than the
expansion rate even for a final temperature $T_f \approx 0$, and the
long wavelength modes will be strongly supercooled.

The purpose of this brief review is to introduce the methods of non-equilibrium
quantum statistical mechanics to study the dynamics of phase transitions,
and apply them to relevant questions in cosmological phase transitions.

Among the topics that we study here are: the ``rolling'' equations and
the corrections introduced by quantum fluctuations, the process of
phase separation via formation and growth  of correlated domains, the
effective equations of motion in Friedmann- Robertson- Walker cosmologies,
including quantum and thermal corrections. We also study the proposal of
the formation of chirally disoriented domains, which may have experimental
relevance in $P-\bar{P}$ and heavy-ion collisions.

\section{\bf Statistical Mechanics out of equilibrium:}

The generalization of statistical mechanics techniques to
the description of non-equilibrium processes in quantum field
theory, have been available for a long
time\cite{schwinger}-\cite{mills} but somehow had
not yet been accepted as an integral part of the available
tools to study field theory in extreme environments. We thus
begin by presenting
 a somewhat pedagogical introduction to the subject for the
non-practioner.

The non-equilibrium description of a system is determined by
the time evolution of the density matrix that describes the
system. This time evolution (in the Schr\"odinger picture) is
determined by the quantum Liouville equation:
\begin{equation}
i\hbar \frac{\partial \rho(t)}{\partial t} = \left[H(t),
\rho(t)\right] \label{liouville1}
\end{equation}

A non-equilibrium situation arises whenever the Hamiltonian
does not commute with the density matrix. Here we allow for
an explicit time dependent Hamiltonian (for example
if the system is in an external time dependent background).

The formal solution of the Liouville equation is:
\begin{equation}
\hat{\rho}(t) = U(t,t_0)\hat{\rho}(t_0)U^{-1}(t,t_0) \label{timedesmat}
\end{equation}
with $\rho(t_0)$ the density matrix at some initial time
$t_0$
that determines the initial condition for the  evolution, and
with $U(t,t_0)$ the time evolution operator.

In most cases of interest the initial density
matrix is either thermal or a pure state corresponding to the
ground state of some initial Hamiltonian. In both cases the
initial density matrix is
\begin{equation}
\rho(t_0) = e^{-\beta H_i}
\end{equation}
the ground state of $H_i$ can be projected out by taking the
limit $\beta \rightarrow \infty$. Non-equilibrium will thus
arise whenever the evolution Hamiltonian does not commute with
the initial Hamiltonian $H_i$. It is convenient to introduce then
a time dependent Hamiltonian $H(t)$ such that $H(t)=H_i$ for
$-\infty \leq t \leq t_0$ and $H(t)=H_{evol}(t)$ for $t > t_0$.
with $H_{evol}(t)$ the evolution Hamiltonian that determines the
dynamics. This corresponds to the assumption that the system
has been in equilibrium up to $t_0$ and will evolve out of
equilibrium thereafter. Real time {\it equilibrium} correlation
functions can be obtained if $H$ is constant in time for all
times.

The expectation
value of any operator is thus
\begin{equation}
\langle {\cal{O}} \rangle(t) = Tr \left[ e^{-\beta_i H_i} U^{-1}(t)
{\cal{O}}U(t)\right]/ Tr \left[ e^{- \beta_i H_i} \right]
\label{expecvalue}
\end{equation}
This expression may be written in a more illuminating form
by choosing an arbitrary time $T <0$ for which
$U(T) = \exp[-iTH_i]$ then we may write
$\exp[-\beta_i H_i] = \exp[-iH_i(T-i\beta_i -T)] = U(T-
i\beta_i,T)$.
Inserting in the trace $U^{-1}(T)U(T)=1$,
commuting $U^{-1}(T)$ with $\hat{\rho}_i$ and using the
composition property of the evolution operator, we may write
(\ref{expecvalue}) as
\begin{equation}
\langle {\cal{O}}\rangle(t) = Tr  \left[U(T-i\beta_i,t) {\cal{O}}
U(t,T)\right]/ Tr  \left[
U(T-i\beta_i,T) \right]\label{trace}
\end{equation}
The numerator of the above expression has a simple meaning:
start at time $T<0$, evolve to time $t$, insert the operator
$\cal{O}$ and evolve backwards in time from $t$ to $T<0$,
and along the negative imaginary axis from $T$ to $T-
i\beta_i$.
The denominator, just evolves along the
negative imaginary axis from $T$ to $T-i\beta_i$. The
contour in
the numerator may be extended to an arbitrary large positive
time
$T'$ by inserting $U(t,T')U(T',t)=1$ to the left of
$\cal{O}$ in
(\ref{trace}), thus becoming
\begin{equation}
\langle {\cal{O}}\rangle(t) = Tr  \left[U(T-
i\beta_i,T)U(T,T')U(T',t){\cal{O}}U(t,T) \right]
/Tr \left[ U(T-i\beta_i,T) \right]
\end{equation}
The numerator now represents the process of evolving from
$T<0$
to $t$, inserting the operator $\cal{O}$, evolving further
to
$T'$, and backwards from $T'$ to $T$ and down the negative
imaginary axis to $T-i\beta_i$. This process is depicted in
the
contour of figure (1). Eventually we take $T \rightarrow -
\infty
\; ; \; T' \rightarrow \infty$. It is straightforward to
generalize
to  real time correlation functions of Heisenberg picture
operators.

This formalism allows us also to study the general case in
which both the mass and the coupling depend on time. The
insertion of the operator ${\cal{O}}$, may be achieved as
usual by introducing currents and taking variational
derivatives with respect to them.

Because the time evolution operators have
the interaction terms in them, and we would like to generate
a perturbative expansion and Feynman diagrams, it is
convenient to introduce source terms for {\it all} the time
evolution operators in the above trace. Thus we are led to
consider the following generating functional
\begin{equation}
Z[J^+, J^-, J^{\beta}]= Tr \left[ U(T-i\beta_i,T;J^{\beta})
U(T,T';J^-)U(T',T;J^+) \right]\label{generatingfunctional}
\end{equation}
The denominator in (\ref{trace}) is simply $Z[0,0,0]$ and
may be obtained in a series expansion in the interaction by
considering $Z[0,0,J^{\beta}]$.
By inserting a complete set of field eigenstates between the
time evolution operators, finally, the generating functional
$Z[J^+,J^-,J^{\beta}]$ may be written as
\begin{eqnarray}
Z[J^+,J^-,J^{\beta}] & = & \int D \Phi D \Phi_1 D \Phi_2
\int {\cal{D}}\Phi^+ {\cal{D}}\Phi^-
{\cal{D}}\Phi^{\beta}\; e^{i\int_T^{T'}\left\{{\cal{L}}[\Phi^+,
J^+]-
{\cal{L}}[\Phi^-,J^-]\right\}}\times   \nonumber\\
                     &   &  e^{i\int_T^{T-
i\beta_i}{\cal{L}}[\Phi^{\beta}, J^{\beta}]}
\label{pathint}
\end{eqnarray}
with the boundary conditions $\Phi^+(T)=\Phi^{\beta}(T-
i\beta_i)=\Phi \;
; \; \Phi^+(T')=\Phi^-(T')=\Phi_2 \; ; \; \Phi^-
(T)=\Phi^{\beta}(T)=
\Phi_1$.
This may be recognized as a path integral along the contour
in complex time shown in Figure (1).
As usual the path integrals over the quadratic forms may be
evaluated and one obtains the final result for the partition
function
\begin{eqnarray}
& & Z[J^+,J^-, J^{\beta}] = \exp{\left\{i\int_{T}^{T'}dt
\left[{\cal{L}}_{int}(-i\delta/\delta J^+)-
{\cal{L}}_{int}(i\delta/\delta
J^-)\right] \right \}} \times \nonumber \\
& &
\exp{\left\{i\int_{T}^{T-
i\beta_i}dt{\cal{L}}_{int}(-i\delta/\delta J^{\beta})
\right\}} \exp{\left\{\frac{i}{2}\int_c
dt_1\int_c dt_2 J_c(t_1)J_c(t_2)G_c (t_1,t_2) \right\}}
 \label{partitionfunction}
\end{eqnarray}

Where $J_c$ stands for the currents on the contour as shown
in figure (1), $G_c$ are the Green's functions on the
contour\cite{niemisemenoff}, and again,
the spatial arguments were suppressed.

In the limit $T \rightarrow \infty$, the contributions from
the terms in which one of the currents is $J^+$ or $J^-$ and
the other is a $J^{\beta}$ vanish when computing correlation
functions in which the external legs are at finite {\it real
time}, as a consequence of the Riemann-Lebesgue lemma. For
this {\it real time} correlation functions, there is no
contribution from the $J^{\beta}$ terms that cancel between
numerator and denominator. Finite temperature enters through
the boundary conditions on the Green's functions (see
below). For the calculation of finite {\it real time}
correlation functions, the generating functional simplifies
to\cite{calzetta,calzettahu}
\begin{eqnarray}
Z[J^+,J^-] & = & \exp{\left\{i\int_{T}^{T'}dt\left[
{\cal{L}}_{int}(-i\delta/\delta J^+)-
{\cal{L}}_{int}(i\delta/\delta
J^-)\right] \right \}} \times \nonumber \\
           &   & \exp{\left\{\frac{i}{2}\int_T^{T'}
dt_1\int_T^{T'} dt_2 J_a(t_1)J_b(t_2)G_{ab} (t_1,t_2)
\right\}}
 \label{generatingfunction}
\end{eqnarray}
with $a,b = +,-$.

  The Green's functions that enter in the integrals along
the  contours in equations (\ref{partitionfunction},
\ref{generatingfunction})
  are given by (see above references)
\begin{eqnarray}
G^{++}(\vec{r}_1,t_1;\vec{r}_2,t_2)  & = &
G^{>}(\vec{r}_1,t_1;\vec{r}_2,t_2)\Theta(t_1-t_2) +
G^{<}(\vec{r}_1,t_1;\vec{r}_2,t_2)\Theta(t_2-t_1)
\label{timeordered}\\
G^{--}(\vec{r}_1,t_1;\vec{r}_2,t_2)  & = &
G^{>}(\vec{r}_1,t_1;\vec{r}_2,t_2)\Theta(t_2-t_1) +
G^{<}(\vec{r}_1,t_1;\vec{r}_2,t_2)\Theta(t_1-t_2)
\label{antitimeordered} \\
G^{+-}(\vec{r}_1,t_1;\vec{r}_2,t_2)  & = & -
G^{<}(\vec{r}_1,t_1;\vec{r}_2,t_2) \label{plusminus}\\
G^{-+}(\vec{r}_1,t_1;\vec{r}_2,t_2)  & = & -
G^{>}(\vec{r}_1,t_1;\vec{r}_2,t_2) = -
G^{<}(\vec{r}_2,t_2;\vec{r}_1,t_1)
\label{minusplus}\\
G^{>}(\vec{r}_1,t_1;\vec{r}_2,t_2)   & = &
\langle\Phi(\vec{r}_1,t_1)\Phi(\vec{r}_2,t_2)\rangle
\label{greater} \\
G^{<}(\vec{r}_1,T;\vec{r}_2,t_2)     & = &
G^{>}(\vec{r}_1,T-i\beta_i;\vec{r}_2,t_2)
\label{periodicity}
\end{eqnarray}

The condition(\ref{periodicity}) is recognized as the
periodicity condition in imaginary time and is a result of
considering an equilibrium situation for $t<0$. The
functions $G^{>}, \; G^{<}$ are homogeneous solutions of the
quadratic form, with appropriate boundary conditions and
will be constructed explicitly below.

There are many clear articles in the literature
using this techniques to study real time correlation
functions\cite{calzetta}-\cite{paz}.

One of our goals is to study the formation and growth of domains and
the time evolution of the correlation functions. In
particular, one relevant quantity of interest is the {\it equal time}
correlation function
\begin{eqnarray}
S(\vec{r};t) & = &
\langle\Phi(\vec{r},t)\Phi(\vec{0},t)\rangle
\label{equaltimecorre} \\
S(\vec{r};t) & = & \int \frac{d^3 k}{(2\pi)^3}\;e^{i\vec{k}
\cdot \vec{r}} \; S(\vec{k};t) \label{strucfacto} \\
S(\vec{k};t) & = & \langle\Phi_{\vec{k}}(t)\Phi_{-
\vec{k}}(t)\rangle =
-iG^{++}_{\vec{k}}(t;t)\label{strufack1}
\end{eqnarray}
where we have performed the Fourier transform in the spatial
coordinates (there still is spatial translational and
rotational invariance). Notice that at equal times, all the
Green's functions are equal, and we may compute any of them.

Clearly in an equilibrium situation this equal time correlation function
will be time independent, and will only measure the
{\it static correlations}. In the present case, however, there is a non
trivial time evolution arising from the departure from equilibrium of the
initial state. This correlation function will measure the correlations in
space, and their time dependence.

The function $G^{>}(\vec{r}_1,t_1;\vec{r}_2,t_2)$ will be constructed
explicitly
below for particular cases, in terms of the homogeneous solutions of the
operator
of quadratic fluctuations.

\section{Rolling with fluctuations}

Inflationary cosmological models provide very appealing
scenarios
to describe the early evolution of our
universe.
Since the original model proposed by Guth\cite{guth1},
several
alternative scenarios have been proposed to overcome some of
the
difficulties with the original proposal.

Among them, the new inflationary
model\cite{guth2}-\cite{linde2}
is perhaps one of the most attractive. The essential
ingredient
in the new inflationary model is a scalar field that
undergoes a
second order phase transition from a high temperature
symmetric phase
to a low temperature broken symmetry phase. The expectation
value (or
thermal average) of the scalar field $\phi$ serves as the
order parameter.
Initially at high temperatures, the scalar field is assumed
to be in
thermal equilibrium and  $\phi \approx 0$.  The usual field-
theoretic tool
to study the phase transition is the effective
potential\cite{kirzhnits}-\cite{sweinberg}. At high  temperatures,
the global minimum of the effective
potential is at $\phi =0$, whereas at low temperatures there
are two
degenerate minima.

The behavior of the phase transition in the new inflationary
model is the
following: as the universe cools down the expectation value
of the scalar
field
remains close to zero until the temperature becomes smaller
than the
critical temperature. Below the critical temperature, when
the effective
potential develops degenerate minima away from the origin,
the scalar
field begins to ``roll down the potential hill''. In the new
inflationary
scenario, the effective potential below the critical
temperature is
extremely flat near the maximum, and the scalar field
remains near the
origin, i.e. the false vacuum for a very long
time and eventually rolls down the hill very slowly.
This scenario thus receives  the name of ``slow rollover''.
During this stage, the energy density of
the universe is dominated by the constant vacuum energy
density $V_{eff}(\phi=0)$, and the universe evolves rapidly into a de Sitter
space (see for
example the reviews by Kolb and Turner\cite{kolb}, Linde\cite{linde}
and Brandenberger\cite{brandenberger}). Perhaps the most remarkable
consequence of the new
inflationary scenario and the slow rollover transition is
that they provide a calculational framework for the prediction
of density fluctuations\cite{starobinsky}. The coupling
constant in the
typical zero temperature potentials must be fine tuned to a
very small value to reproduce the observed limits on  density
fluctuations\cite{kolb,linde}.

This picture of the slow rollover scenario, is based on the
{\it static}
effective potential whose use we already criticized in the previous
section.

Guth and Pi\cite{guthpi}, performed a thorough analysis of
the effects
of quantum fluctuations on the time evolution. These authors
analyzed
the situation below the critical temperature by treating the
potential
near the origin as an {\it inverted harmonic oscillator}.
They recognized that the
instabilities associated with these upside-down oscillators
lead to an
exponential growth of the quantum fluctuations at long times
and to a
classical description of the probability distribution
function. Guth
and Pi also recognized that the {\it static} effective
potential is not
appropriate to describe the dynamics, that must be treated
as a time
dependent process.

The purpose of this section is to study the influence of the
quantum and thermal fluctuations
on the ``slow-rollover'' picture. In particular the effect of the
growth of correlations on the evolution of the expectation value of
the inflaton field.

In order to understand the dynamics of the phase transition and
the physics of the instabilities mentioned above, let us consider
the situation in which for time $t<0$ the system is in {\it
equilibrium} at an initial temperature $T_i > T_c$ where
$T_c$ is the critical temperature.
At time $t = 0$ the system is  rapidly ``quenched''
to a final temperature below the critical temperature
$T_f < T_c$ and evolves thereafter out of equilibrium.
We will postpone to a later section the study of the evolution equations
in an expanding cosmology.

Here we just assume that such a mechanism takes place and
simplify the situation by introducing a Hamiltonian with a
{\it time dependent mass term} to describe this situation
\begin{eqnarray}
  H(t) & = & \int_{\Omega} d^3x \left\{
\frac{1}{2}\Pi^2(x)+\frac{1}{2}(\vec{\nabla}\Phi(x))^2+\frac
{1}{2}m^2(t)
\Phi^2(x)+\frac{\lambda}{4!}\Phi^4(x) \right \}
\label{timedepham} \\
m^2(t) & = & m^2 \, \Theta(-t)  -\mu^2\, \Theta(t)
\label{massoft}
\end{eqnarray}
where both $m^2$ and $\mu^2$ are positive. We assume that
for all times $t <0$ there is thermal equilibrium at
temperature $T_i$, and the system is described by the
density matrix
\begin{eqnarray}
\hat{\rho}_i  & = & e^{-\beta_i H_i} \label{initialdesmat}\\
          H_i & = & H(t<0) \label{initialham}
\end{eqnarray}

An alternative and equally valid interpretation (and the one
that
we like best) is that the initial
condition  being considered here is that of a system in
equilibrium in the
symmetric phase, and evolved in time with a Hamiltonian that
allows for
broken symmetry ground states, i.e.  the Hamiltonian (\ref
{timedepham}, \ref{massoft}) for $t>0$.

Using the tools of non-equilibrium statistical mechanics introduced in the
previous section we are now in condition to obtain the evolution equations
for the average of the scalar field in the case when the
potential is
suddenly changed. To account for a sudden change in
temperature
from above to below the critical
temperature as described by the model Hamiltonian
(\ref{timedepham}) with (\ref{massoft}).
 For this purpose we use the tadpole method\cite{sweinberg},
and write
\begin{equation}
\Phi^{\pm}(\vec{x},t) = \phi(t) + \Psi^{\pm}(\vec{x},t)
\label{expecvaluepsi}
\end{equation}
Where, again, the $\pm$ refer to the branches for forward
and backward time propagation. The reason for shifting both
($\pm$) fields by the {\it same} classical configuration, is
that $\phi(t)$ enters in the time evolution operator as a
background c-number variable, and time evolution forward and
backwards are now considered in this background.

The  evolution equations are obtained with the
tadpole method by expanding the Lagrangian around $\phi(t)$
and considering the {\it linear}, cubic, quartic, and higher
order terms in $\Psi^{\pm}$ as perturbations and requiring
that
\[ <\Psi^{\pm}(\vec{x},t) >= 0 .\]

It is a straightforward exercise to see that this is
equivalent to  one-loop to extremizing
the one-loop effective action in which the determinant (in
the log det)
incorporates the boundary condition of equilibrium at time
$t<0$ at the initial temperature.

To one loop we find the  equation of motion
\begin{equation}
\frac{d^2\phi(t)}{dt^2}+m^2(t)\phi(t)+\frac{\lambda}{6}
\phi^3(t)+ \frac{\lambda}{2}\phi(t)\int \frac{d^3
k}{(2\pi)^3}(-i) G_k(t,t) = 0 \label{eqofmotion}
\end{equation}
with $G_k(t,t)=G_k^{<}(t,t)=G_k^{>}(t,t)$ is the spatial
Fourier transform of the equal-time Green's function.

At this point, we would like to remind the reader that

\[-iG_k(t,t) =<\Psi^+_{\vec{k}}(t) \Psi^+_{-\vec{k}}(t)>\]

is a {\it positive definite quantity} (because the field
$\Psi$ is real) and as we argued before
(and will be seen explicitly shortly) this Green's function
grows in time because of the instabilities associated with
the phase transition and domain growth\cite{weinbergwu,guthpi}.

These Green's functions are constructed out of the
homogeneous solutions to the operator of quadratic
fluctuations
\begin{eqnarray}
\left[\frac{d^2}{dt^2} + \vec{k}^2 +
M^2(t)\right]{\cal{U}}_k^{\pm}(t) & = & 0 \label{homogene}\\
 M^2(t)  =  (m^2+\frac{\lambda}{2}
\phi^2_i)\Theta(-t)            & + &
[-\mu^2+\frac{\lambda}{2}\phi^2(t)]\Theta(t)
\label{bigmassoft}
\end{eqnarray}

The boundary conditions on the homogeneous solutions are
\begin{eqnarray}
{\cal{U}}_k^{\pm}(t<0) & = & e^{\mp i
\omega_{<}(k)t}\label{boundarycondi} \\
\omega_{<}(k)          & = &
\left[\vec{k}^2+m^2+\frac{\lambda}{2}\phi^2_i\right]^{\frac{
1}{2}}\label{omegamin}
\end{eqnarray}
where $\phi_i$ is the value of the classical field at time
$t<0$ and is the initial boundary condition on the equation
of motion. Truly speaking, starting in a fully symmetric
phase will force $\phi_i =0$, and the time evolution will maintain
this value, therefore we admit a small explicit symmetry
breaking field in the initial density matrix to allow for a
small $\phi_i$. The introduction of this initial condition
seems artificial since we are studying the situation of
cooling down from the symmetric phase.

 However, we recognize that  the phase transition from the symmetric phase
 occurs via formation of domains (in the case of a discrete symmetry)
inside which the order parameter acquires non-zero values. The domains
will have the same probability for either value of the field and the
volume average of the field will remain zero. These domains
will grow in time, this is the phenomenon of phase
separation and spinodal decomposition
familiar in condensed matter physics. Our evolution equations presumably
will apply to the coarse grained average of the scalar field inside each
of these domains. This average will only depend on time. Thus, we interpret
$\phi_i$ as corresponding to the coarse grained average
of the field in each of these domains.
The question of initial conditions on the scalar field is
also present (but usually overlooked) in the slow-rollover
scenarios but as we will see later, it plays a fundamental
role in the description of the evolution.

The identification of the initial value $\phi_i$ with the
average of the field in each domain is certainly a plausibility argument to
justify an initially small asymmetry in the scalar field which is necesary
for the further evolution of the field, and is consistent with the usual
assumption within the slow rollover scenario.

The boundary conditions on the mode functions
${\cal{U}}_k^{\pm}(t)$ correspond to  ``vacuum'' boundary
conditions of positive and negative frequency modes
(particles and antiparticles) for $t<0$.

Finite temperature enters through the periodicity conditions
(\ref{periodicity}) and the Green's functions are
\begin{eqnarray}
G^{>}_k(t,t') & = & \frac{i}{2\omega_<(k)} \frac{1}{1-e^{-
\beta_i \omega_<(k)}}\left[{\cal{U}}_k^+(t) {\cal{U}}_k^-
(t')+ e^{-\beta_i\omega_<(k)} {\cal{U}}_k^-(t)
{\cal{U}}_k^+(t') \right] \label{finalgre} \\
G^{<}_k(t,t') & = & G^{>}(t',t)
\end{eqnarray}

Summarizing, the effective equations of motion to one loop
that determine the time evolution of the scalar field are
\begin{eqnarray}
\frac{d^2\phi(t)}{dt^2}+m^2(t)\phi(t)  +
\frac{\lambda}{6} \phi^3(t) & + &
               \frac{\lambda}{2}\phi(t) \int \frac{d^3
k}{(2\pi)^3} \frac{{\cal{U}}_k^+(t) {\cal{U}}_k^-(t)}{2\omega_<(k)}
\mbox{coth}\left[\frac{\beta_i \omega_<(k)}{2}\right] =  0
\label{finaleqofmotion1} \\
\left[\frac{d^2}{dt^2} + \vec{k}^2 +  M^2(t)\right]{\cal{U}}_k^{\pm}
          & = & 0 \label{finaleqofmotion2}
\end{eqnarray}
with (\ref{bigmassoft}) , (\ref{boundarycondi}).

This set of equations is too complicated to attempt an  analytic
solution, we will study this system numerically shortly.

However, before studying numerically these equations, one recognizes that
there are several features of this set of equations
that reveal the basic physical aspects of the dynamics of the scalar field.

{\bf i)}: The effective evolution equations are {\bf real}.
The mode functions ${\cal{U}}_k^{\pm}(t)$ are complex conjugate of  each other
as may be seen from the reality of the equations,
and the boundary conditions (\ref{boundarycondi}).
This  situation must be contrasted with the expression for the
effective potential for the {\it analytically continued modes}.

{\bf ii)}: Consider the situation in which the initial
configuration of the classical field is near the origin $\phi_i \approx 0$,
for $t>0$, the modes for which $\vec{k}^2 < (k_{max})^2 \; ; \;
(k_{max})^2 = \mu^2-\frac{\lambda}{2}\phi_i^2$ are {\it unstable}.

In particular, for early times ($t>0$), when $\phi_i \approx 0$,
these unstable modes behave approximately as
\begin{eqnarray}
{\cal{U}}_k^+(t) & = & A_k \, e^{W_k t}+B_k \, e^{-W_k t}
\label{unstable1}\\
{\cal{U}}_k^-(t) & = & ({\cal{U}}_k^+(t))^{*}
\label{unstable2}\\
A_k              & = & \frac{1}{2}\left[1-i\frac{\omega_{<}(k)}
{W_k} \right] \; ; \; B_k = A_k^* \label{abofk} \\
W_k              & = & \left[\mu^2-\frac{\lambda}{2}\phi_i^2
- \vec{k}^2 \right]^{\frac{1}{2}} \label{bigo}
\end{eqnarray}
Then the early time behavior of $-iG_k(t,t)$ is given by
\begin{equation}
-iG_k(t,t) \approx \frac{1}{2\omega_<(k)}
\left[1+\frac{\mu^2+m^2}
{\mu^2-\frac{\lambda}{2} \phi_i^2-k^2}[\cosh(2W_kt)-1]\right]
\coth[\beta_i\omega_<(k)/2]
\label{earlytimegreen}
\end{equation}
This early time behavior coincides with that of the Green's function
of Guth and Pi\cite{guthpi} and Weinberg and Wu\cite{weinbergwu} for the
inverted harmonic oscillators when our initial
state (density matrix) is taken into account.  An important point to notice
is that for early times all the wavevectors in the unstable band grow with
{\em the same rate}.

Our evolution equations, however, permit us to go beyond the
early time behavior and to incorporate the non-linearities that
will eventually shut off the instabilities.

These early-stage instabilities and subsequent growth of fluctuations
and correlations, are the hallmark of the process of phase separation,
and precisely the instabilities that trigger the phase transition.

It is clear from the above equations of evolution, that the
description in terms
of inverted oscillators will only be valid at very early times.

At early times, the {\it stable} modes for which $\vec{k}^2>(k_{max})^2$
are obtained from (\ref{unstable1}) , (\ref{unstable2}) , (\ref{abofk})
by the  analytic continuation
$ W_k \rightarrow -i \omega_{>}(k)= \left[ \vec{k}^2-\mu^2+
\frac{\lambda}{2}\phi_i^2 \right]^{\frac{1}{2}} $.

For $t<0$, ${\cal{U}}_k^+(t){\cal{U}}_k^-(t)=1$ and one
obtains
the usual result for the evolution equation
\[ \frac{d^2 \phi(t)}{dt^2}+\frac{dV_{eff}(\phi)}{d\phi} =0 \]
with $V_{eff}(\phi)$ the finite temperature effective
potential but for $t<0$ there are no unstable modes.

It becomes clear, however, that for $t>0$ there are no {\it static}
solutions to the evolution equations for $\phi(t) \neq 0$.

{\bf iii)} {\bf Coarsening}: as the classical expectation
value
$\phi(t)$ ``rolls down'' the potential hill, $\phi^2(t)$
increases and $(k_{max}(t))^2= \mu^2-
\frac{\lambda}{2}\,\phi^2(t)$
{\it decreases}, and only the very
long-wavelength modes remain unstable, until for a
particular
time  $t_s\; ; \; (k_{max}(t_s))^2=0$. This occurs when
$\phi^2(t_s) = 2\mu^2/\lambda$, this is the inflexion point
of the tree level potential. In statistical mechanics this point is
known as the ``classical spinodal point'' and $t_s$ as the
``spinodal time''\cite{langer,guntonmiguel}. When the
classical field reaches the spinodal point, all instabilities shut-
off. {F}rom this point on, the dynamics is oscillatory and this
period is identified with the ``reheating'' stage in cosmological
scenarios\cite{linde,brandenberger}.

It is clear from the above equations of evolution, that the
description in terms of inverted oscillators will only be valid at
small positive times, as eventually the unstable growth will shut-off.

The value of the spinodal time depends on the
initial conditions of $\phi(t)$. If the initial value $\phi_i$ is
very near the classical spinodal point, $t_s$ will be relatively
small and there will not be enough time for the unstable
modes to grow too much.
In this case, the one-loop corrections for small coupling
constant will remain perturbatively small.
On the other hand, however, if $\phi_i \approx 0$, and the
initial velocity is small, it will take a very long time to reach the
classical spinodal point. In this case the unstable modes
{\bf may grow dramatically making the one-loop corrections
non-negligible even for
small coupling}. These initial conditions of small initial field and
velocity are precisely the ``slow rollover'' conditions that
are of interest in cosmological scenarios of ``new
inflation''.

The  renormalization aspects are too technical for this brief review
and we refer the reader to our original article\cite{boyveg} for
details.

 As mentioned previously within the context of coarsening, when
 the initial value of the scalar field $\phi_i \approx 0$, and
 the initial temporal derivative is small, the scalar field slowly
 rolls down the potential hill. But during the time while the scalar
 field remains smaller than the ``spinodal'' value, the unstable
 modes grow and the one-loop contribution grows consequently. For a
 ``slow rollover'' condition, the field remains very small
($\phi^2(t) \ll 2\mu^2/\lambda$) for a long
 time, and during this time the unstable modes grow exponentially.
 The stable modes, on the other hand give an oscillatory contribution
 which is bound in time, and for weak coupling remains
perturbatively small at all times.

 Then for a ``slow rollover'' situation and for
 weak coupling, only the unstable modes will yield to an important
 contribution to the one-loop correction. Thus, in the evolution equation
 for the scalar field, we will keep only the integral over the {\it
 unstable modes} in the one loop correction.

Phenomenologically the coupling constant in these models is
bound by the spectrum of density fluctuations to be within
the range $\lambda_R \approx 10^{-12} - 10^{-
14}$\cite{linde}. The the stable modes will
\underline{always} give a \underline{perturbative}
contribution, whereas the unstable modes grow exponentially
in time thus raising the possibility of giving a non-
negligible contribution.

With the purpose of numerical analysis of the
equations
 of motion, it proves convenient to introduce the following
 dimensionless variables
 \begin{eqnarray}
 \tau      & = & \mu_R t \; ; \;  q   =  k/\mu_R \\
 \eta^2(t) & = & \frac{\lambda_R}{6\mu_R^2}\phi^2(t) \; ; \;
L^2 =
 \frac{m_R^2 + \frac{1}{2}\lambda_R \phi_i^2}{\mu^2_R}
 \end{eqnarray}
 and to account for the change from the initial temperature
to the final
 temperature ($T_i>T_c \; ; \; T_f<T_c$) we
parametrize
 \begin{eqnarray}
 m^2   & = & \mu_R(0)\left[\frac{T_i^2}{T_c^2} -1 \right]
\label{abovecrit} \\
 \mu_R & = & \mu_R(0)\left[1-\frac{T_f^2}{T_c^2} \right]
\label{belowcrit}
 \end{eqnarray}
 where the subscripts (R) stand for  renormalized
quantities, and
 $-\mu_R(0)$ is the renormalized zero temperature ``negative
mass squared''
 and $T^2_c = 24 \mu^2_R(0)/\lambda_R$.  Furthermore,
because $(k_{max}(t))^2
 \leq \mu^2_R$ and $T_i > T_c$, for the unstable modes $T_i
\gg (k_{max}(t))$
 and we can take the high temperature limit
$\mbox{coth}[\beta_i
 \omega_<(k)/2] \approx 2 T_i/\omega_<(k)$.  Finally the
effective equations of evolution for $t > 0$, keeping in the one-loop
contribution only the unstable modes  as explained above
($q^2 <(q_{max}(\tau))^2$) become, after using $\omega_<^2 = \mu_R^2(q^2+L^2)$
 \begin{eqnarray}
 \frac{d^2}{d\tau^2}\eta(\tau)-\eta(\tau) +  \eta^3(\tau)  & +
                                            & g\eta(\tau)
\int_0^{(q_{max}(\tau))}q^2\frac{{\cal{U}}^+_q(\tau){\cal{U}}^-_q(\tau)}
  {q^2+L^2}dq = 0 \label{eqofmotunst}\\
  \left[\frac{d^2}{d\tau^2}+ q^2-
(q_{max}(t))^2\right]{\cal{U}}^{\pm}_q(\tau)
                                         & = & 0
\label{unstamod} \\
                       (q_{max}(\tau))^2 & = & 1-3\eta^2(\tau)
  \label{qmaxoft} \\
                                      g  & = & \frac{\sqrt{6\lambda_R}}
  {2\pi^2}\frac{T_i}{T_c \left[1-\frac{T^2_f}{T^2_c}\right]}
\label{coupling}
  \end{eqnarray}

For $T_i \geq T_c$ and $T_f \ll T_c$ the coupling (\ref{coupling}) is bound
within the range $g \approx 10^{-7}-10^{-8}$.
 The dependence of the coupling with the temperature  reflects the
 fact that at higher temperatures the fluctuations are enhanced.
 It is then clear, that the contribution from the stable modes
 is {\it always perturbatively small}, and only the unstable modes may
 introduce important corrections if they are allowed to grow for a long  time.

 From (\ref{eqofmotunst}) we see that the quantum corrections act as a
 {\it positive dynamical renormalization} of the ``negative mass'' term
 that drives  the rolling down dynamics. It is then clear, that the
 quantum corrections tend to {\it slow down the evolution}.

In particular,
 if the initial value $\eta(0)$ is very small, the unstable modes grow for
 a long time before $\eta(\tau)$ reaches the spinodal point
$\eta(\tau_s) = 1/\sqrt{3}$ at which point the instabilities shut
off. The time $\tau_s$ when this point is attained is called the
spinodal time.
If  this is the case, the quantum corrections introduce substantial
modifications to the classical equations of motion, thus becoming non-
perturbative. If $\eta(0)$ is closer to the classical spinodal point, the
unstable modes do not have time to grow dramatically and the quantum
corrections are perturbatively small.

 Thus we conclude that the initial conditions on the field
determine whether or not the quantum corrections are perturbatively
small.

 Although the system of equations (\ref{eqofmotunst}, \ref{unstamod}),
 are coupled, non-linear and integro-differential, they may be integrated
 numerically. Figures (2,3,4) depict the solutions for the classical
 (solid lines) and quantum (dashed lines) evolution (a),
 the quantum correction (b), i.e. the fourth term in (\ref{eqofmotunst})
 {\it including the coupling} $g$, the classical (solid lines) and
 quantum (dashed lines) velocities
$\frac{d\eta(\tau)}{d\tau}$ (c) and $(q_{max}(\tau))^2$ (d).
For the numerical integration, we have
 chosen $L^2 =1$, the results are only weakly dependent on $L$, and
 taken $g=10^{-7}$, we have varied the initial condition $\eta(0)$ but
 used $\frac{d\eta(\tau)}{d\tau}|_{\tau=0} = 0$.

 We recall, from a previous discussion that $\eta(\tau)$ should be
 identified with the average of the field within a domain. We are
 considering the situation in which this average is very small, according
 to the usual slow-rollover hypothesis, and for  which the instabilities
 are stronger.

 In figure (2.a)
 $\eta(0) = 2.3 \times 10^{-5}$ we begin to see that the
quantum  corrections become important at $\tau \approx 10$ and
slow down the dynamics. By the time that the {\it classical}
evolution  reaches the minimum of the classical potential at $\eta
=1$, the quantum evolution has just reached the classical spinodal $\eta =
1/\sqrt{3}$.
 We see in figure (2.b) that the quantum correction becomes large enough
 to change the sign of the ``mass term'', the field continues its
 evolution towards larger values, however, because the velocity is
 different from zero, attaining a maximum (figure (2.c)) around the
 time when the quantum correction attains its maximum. As $\eta$ gets
 closer to the classical spinodal point, the unstability shuts off as is
 clearly seen in figure (2.d) and the quantum correction arising from the
 unstable modes become perturbatively small. From the spinodal point
 onwards, the field evolves towards the minimum and begins to oscillate
 around it, the quantum correction will be perturbatively small, as all
 the instabilities had shut-off. Higher order corrections, will introduce
 a damping term as quanta may decay in elementary excitations of the true
 vacuum.

 Figures (3.a-d) show a more marked behavior for $\eta(0)= 2.27 \times
 10^{-5}\; ; \; \frac{d\eta (0)}{d\tau}=0$, notice that the classical
 evolution of the field has reached beyond the minimum of the potential
 at the time when the quantum evolution has just reached the classical
 spinodal point. Figure (3.b) shows that the quantum correction becomes
 larger than $1$, and dramatically slows down the evolution, again
 because the velocity is different from zero (figure (3.c)) the field
 continues to grow. The velocity reaches a maximum and begins to drop.
 Once the field reaches the spinodal again the instabilities shut-off
 (figures (3.b,d)) and from this point the field will continue to evolve
 towards the minimum of the potential but the quantum corrections will
 be perturbatively small.

 Figures (4.a-d) show a dramatic behavior for $\eta(0) = 2.258 \times
 10^{-5} \; ; \; \frac{d\eta (0)}{d\tau} = 0$. The unstable modes have
 enough time to grow so dramatically that the quantum correction (figure
 (4.b)) becomes extremely large $\gg 1$ (figure(4.b)), overwhelming the
 ``negative mass'' term near the origin. The dynamical time dependent
potential, now becomes {\it a minimum} at the origin and the quantum evolution
begins to {\it oscillate} near $\eta = 0$. The contribution of the
unstable modes has become {\it non-perturbative}, and certainly our one-loop
 approximation breaks down.

 As the initial value of the field gets closer to zero, the unstable modes
 grow for a very long time. At this point, we realize, however, that this
 picture cannot be complete. To see this more clearly, consider the case
 in which the initial state
 or density matrix corresponds exactly to the symmetric case. $\eta =0$ is
 necessarily, by symmetry, a fixed point of the equations of motion.
 Beginning from the symmetric state, the field will {\it always remain} at
 the origin and though there will be strong quantum and thermal fluctuations,
 these are symmetric and will sample field configurations with opposite
 values of the field with equal probability.

 In this situation, and according to the picture presented above, one would
 then expect that the unstable modes will grow indefinetely because the
 scalar field does not roll down and will never reach the classical spinodal
 point thus shutting-off the instabilities. What is missing in this picture
 and the resulting equations of motion is a self-consistent treatment of the
 unstable fluctuations, that must necessarily go beyond one-loop. A more
 sophisticated and clearly non-perturbative scheme must be invoked that will
 incorporate coarsening, that is the shift with time of the unstable modes
 towards longer wavelength and the eventual shutting off of the instabilities.

\section{Dynamics of Phase Separation: Domain formation and Growth}

In the previous section we addressed the issue of the ``slow-rollover'',
and we found that the growth of correlations that drives the process of
phase separation (``spinodal decomposition'') and is the hallmark of the phase
transition makes dramatic corrections to the classical evolution of the
expectation value of the scalar field.

Our goal in this section is to study the formation and growth of correlated
domains and
the time evolution of the correlation functions. We now consider the case of
a rapid supercooling from the symmetric phase at $T_i > T_c$ into
the low temperature
phase for $T_f < T_c$ with $\langle \phi \rangle=0$.
This value of the expectation
value is a fixed point of the equations of motion. Thus the phase transition
will occur via the formation and growth of correlated domains,
but statistically
there will be domains in which the average of the scalar field will be positive
and others in which it will be negative, with zero average.

In this situation, as argued in section 2 the relevant quantity of
interest is the {\it equal time} correlation function
\begin{eqnarray}
S(\vec{r};t) & = &
\langle\Phi(\vec{r},t)\Phi(\vec{0},t)\rangle
\label{equaltimecorr} \\
S(\vec{r};t) & = & \int \frac{d^3 k}{(2\pi)^3}\;e^{i\vec{k}
\cdot \vec{r}} \; S(\vec{k};t) \label{strucfac} \\
S(\vec{k};t) & = & \langle\Phi_{\vec{k}}(t)\Phi_{-
\vec{k}}(t)\rangle =
-iG^{++}_{\vec{k}}(t;t)\label{strufack}
\end{eqnarray}

The function $G^{>}_{\vec{k}}(t,t')$ is constructed from the
homogeneous solutions to the operator of quadratic
fluctuations
\begin{eqnarray}
\left[\frac{d^2}{dt^2} + \vec{k}^2 +
m^2(t)\right]{\cal{U}}_k^{\pm} & = & 0 \label{homogeneous}
 \end{eqnarray}
with $m^2(t)$ given by (\ref{bigmassoft}).

The boundary conditions on the homogeneous solutions are
\begin{eqnarray}
{\cal{U}}_k^{\pm}(t<0) & = & e^{\mp i
\omega_{<}(k)t}\label{boundaryconditions} \\
\omega_{<}(k)          & = &
\left[\vec{k}^2+m^2_i\right]^{\frac{1}{2}}\label{omegaminus}
\end{eqnarray}
corresponding to positive frequency (${\cal{U}}_k^{+}(t)$)
(particles) and negative
frequency
(${\cal{U}}_k^{-}(t)$), (antiparticles) respectively.

The solutions are as follows:
i): stable modes ($\vec{k}^2 > m^2_f$)
\begin{eqnarray}
{\cal{U}^+}_k(t) & = & e^{-i\omega_{<}(k) t}\,\Theta(-t)+
\left(a_k e^{-i\omega_{>}(k)t}+b_k
e^{i\omega_{>}(k)t}\right)\Theta(t) \label{stablemodes} \\
{\cal{U}^-}_k(t) & = & \left({\cal{U}^+}_k(t)\right)^{*} \\
\omega_{>}(k)    & = & \sqrt{\vec{k}^2-m^2_f}
\label{omegaplus} \\
a_k              & = &
\frac{1}{2}\left(1+\frac{\omega_{<}(k)}{\omega_{>}(k)}
\right) \\
b_k              & = & \frac{1}{2}\left(1-
\frac{\omega_{<}(k)}{\omega_{>}(k)}\right) \\
\end{eqnarray}

ii): unstable modes ($\vec{k}^2 < m^2_f$)
\begin{eqnarray}
{\cal{U}^+}_k(t) & = & e^{-i\omega_{<}(k) t}\,\Theta(-t)+
\left(A_k e^{W(k)t}+B_k e^{-W(k)t}\right)\Theta(t)
\label{unstablemodes} \\
{\cal{U}^-}_k(t) & = & \left({\cal{U}^+}_k(t)\right)^{*} \\
W(k)             & = & \sqrt{m^2_f-\vec{k}^2} \label{bigW}
\\
A_k              & = & \frac{1}{2}\left(1-
i\frac{\omega_{<}(k)}{W(k)}\right) \; \; ; \; \; B_k =
(A_k)^*  \\
\end{eqnarray}
With these mode functions, and the periodicity condition
(\ref{periodicity}) we find
\begin{eqnarray}
G^{>}_k(t,t') & = & \frac{i}{2\omega_< (k)} \frac{1}{1-
e^{-
\beta_i \omega_< (k)}}\left[{\cal{U}}_k^+(t)
{\cal{U}}_k^-
(t')+ e^{-\beta_i\omega_< (k)} {\cal{U}}_k^-(t)
{\cal{U}}_k^+(t') \right] \label{finalgreenfunc} \\
G^{<}_k(t,t') & = & G^{>}(t',t)
\end{eqnarray}

 The zeroth order equal time Green's function becomes
\begin{equation}
 G^{>}_{\vec{k}}(t;t)= \frac{i}{2 \omega_{<}(k)}
 \coth[\beta_i\omega_{<}(k)/2]
\end{equation}
for $t< 0$, and

\begin{eqnarray}
G^{>}_{\vec{k}}(t;t) & = & \frac{i}{2 \omega_{<}(k)} \{
\left[1+2A_kB_k\left[ \cosh(2W(k)t)-1 \right]\right]
 \Theta(m^2_f-\vec{k}^2)
\nonumber  \\
                       & + &
\left[1+2a_kb_k\left[ \cos(2 \omega_{>}(k)t)-1
\right]\right]
 \Theta(\vec{k}^2-m^2_f)\}
\coth[\beta_i\omega_{<}(k)/2]
\end{eqnarray}
for $t > 0$.

The first term, the contribution of the unstable modes,
reflects the growth of correlations because of the
instabilities and will be the dominant term at long times.

\section{\bf Zeroth order correlations:}

Before proceeding to study the correlations in higher orders
in the coupling
constant, it will prove to be very illuminating to
understand the behavior
of the equal time non-equilibrium correlation functions at
tree-level.
Because we are interested in the  growth of correlations, we
will study only
the contributions of the unstable modes.

 The integral of the equal time correlation function over
all wave
vectors shows the familiar short distance divergences.  From
the above
expression; however, it is clear that these may be removed
by subtracting (and also multiplicatively renormalizing)
this correlation function at $t=0$.
The contribution of the stable modes to the
subtracted and multiplicatively renormalized correlation function is
always bounded in time and
thus uninteresting
for the purpose of understanding the growth of the
fluctuations.

We are thus
led to study {\it only} the contributions of the unstable
modes to the  subtracted and renormalized
correlation function, this contribution is finite and unambiguous.

For this purpose it is convenient to introduce the following
dimensionless quantities
\begin{equation}
\kappa = \frac{k}{m_f}  \; \; ; \; \;  L^2 = \frac{m_i^2}{m_f^2}=
\frac{T^2_i-T^2_c}{T^2_c-T^2_f}
\; \; ; \; \; \tau = m_f t  \; \; ; \; \;  \vec{x} = m_f \vec{r}
\label{dimensionless}
\end{equation}
Furthermore for the unstable modes $\vec{k}^2 < m^2_f$, and
for initial
temperatures  larger than the critical temperature
$T^2_c = 24 \mu^2 / \lambda$,  we can approximate $\coth[\beta_i
\omega_{<}(k)/2] \approx
2 T_i / \omega_{<}(k)$. Then, at tree-level, the
contribution of the
unstable modes to the subtracted structure factor
(\ref{strufack})
$S^{(0)}(k,t)-
S^{(0)}(k,0)=(1/m_f){\cal{S}}^{(0)}(\kappa,\tau)$ becomes
\begin{eqnarray}
{\cal{S}}^{(0)}(\kappa,\tau)   & = &
\left( \frac{24}{\lambda [1-\frac{T^2_f}{T^2_c}]} \right)^{\frac{1}{2}}
\left(\frac{T_i}{T_c} \right)
\frac{1}{2\omega^2_{\kappa}} \left( 1+
\frac{\omega^2_{\kappa}}{W^2_{\kappa}}
\right) \left[ \cosh(2W_{\kappa}\tau)-1 \right]
\label{strucuns} \\
              \omega^2_{\kappa}& = & \kappa^2+L^2 \label{smallomegak} \\
  W_{\kappa}                   & = & 1-\kappa^2 \label{bigomegak}
\end{eqnarray}

To obtain a better idea of the growth of correlations, it is
convenient to
introduce the scaled correlation function
\begin{equation}
{\cal{D}}(x,\tau) = \frac{\lambda}{6m^2_f}\int^{m_f}_0
\frac{k^2 dk}{2\pi^2}\frac{\sin(kr)}{kr}\,[S(k,t)-S(k,0)]
\label{integral}
\end{equation}
The reason for this is that the minimum of the tree level
potential occurs
at $\lambda \Phi^2 /6 m^2_f =1$, and the inflexion
(spinodal) point,
at $\lambda \Phi^2 /2 m^2_f =1$, so that ${\cal{D}}(0,\tau)$
measures the
excursion of the fluctuations to the spinodal point and
beyond as the correlations grow in time.

At large $\tau$ (large times),
the  product $\kappa^2 {\cal{S}}(\kappa,\tau)$ in
(\ref{integral}) has a very sharp peak at
$\kappa_s = 1/ \sqrt{\tau}$. In the region $x < \sqrt{\tau}$
the integral
may be done by the  saddle point approximation and we obtain for
$T_f/T_c \approx 0$ the large time behavior
\begin{eqnarray}
{\cal{D}}(x,\tau) & \approx & {\cal{D}}(0,\tau)
\exp[-\frac{x^2}{8\tau}]
\frac{\sin(x/ \sqrt{\tau})}{(x/ \sqrt{\tau})}
\label{strucfacxtau} \\
{\cal{D}}(0,\tau) & \approx & \left(\frac{\lambda}{12
\pi^3}\right)^
{\frac{1}{2}}\left(\frac{(\frac{T_i}{2 T_c})^3}{[
\frac{T^2_i}{T^2_c}-
1]}\right)\frac{\exp[2\tau]}{\tau^{\frac{3}{2}}}
\label{strucfactau}
\end{eqnarray}

 Restoring dimensions, and recalling that the zero
temperature correlation
 length is $\xi(0) = 1/\sqrt{2} \mu$,
 we find that for $T_f \approx 0$ the amplitude of the
fluctuation inside a
 ``domain'' $\langle \Phi^2(t)\rangle$, and the ``size'' of
a  domain  $\xi_D(t)$ grow as
 \begin{eqnarray}
 \langle \Phi^2(t)\rangle & \approx &  \frac{\exp[\sqrt{2}t/
\xi(0)]}
 {(\sqrt{2}t/ \xi(0))^{\frac{3}{2}}} \label{domainamplitude}
\\
 \xi_D(t)      & \approx & (8\sqrt{2})^{\frac{1}{2}}
 \xi(0)\sqrt{\frac{t}{\xi(0)}}
\label{domainsize}
 \end{eqnarray}

 An important time scale corresponds to the time $\tau_s$ at
which the  fluctuations
 of the field sample beyond the spinodal point. Roughly
speaking when this
 happens, the instabilities should shut-off as the mean
square root
 fluctuation of the field $\sqrt{\langle\Phi^2(t)\rangle}$
is now probing the stable  region.
 This will be seen explicitly below when we study the
evolution  non-perturbatively in the Hartree approximation and
the fluctuations are incorporated self-consistently in the evolution
equations.
 In zeroth order we estimate this time
from the condition
 $3{\cal{D}}(0,t) = 1$, we use
 $\lambda= 10^{-12} \; ; \;  T_i/T_c=2$,
 as representative parameters
 (this value
 of the initial temperature does not have  any particular physical
 meaning and was chosen only as  representative). We find
 \begin{equation}
 \tau_s \approx 10.15 \label{spinodaltime}
 \end{equation}
 or in units of the zero temperature correlation length
 \begin{equation}
 t \approx 14.2 \; \xi(0)
 \end{equation}
 for other values of the parameter $\tau_s$ is found from
the above condition
 on (\ref{strucfactau}).

In reference\cite{boydom} we showed explicitly that the time
evolution cannot be studied in perturbation theory because of the
unstabilities associated with the process of phase separation and
domain growth. These are characterized by exponentially growing
contributions that even for small coupling will grow to become larger
than the zeroth order contributions. Thus we need a non-perturbative
treatment.

{\subsection{\bf Beyond perturbation theory: Hartree approximation}}

It became clear from the analysis of the previous section that perturbation
theory is inadequate to describe the non-equilibrium dynamics of the phase
transition, precisely because of the instabilities and the growth of
correlations. This growth is manifest in the Green's functions that enter in
any perturbative expansion thus invalidating any perturbative approach. Higher
order corrections will have terms that grow exponentially and faster than the
previous term in the expansion. And even for very weakly coupled theories,
the higher order corrections eventually become of the same
order as the lower order terms.

As the correlations and fluctuations grow, field
configurations start sampling the stable region beyond the spinodal point.
This will result in  a slow down in the
growth of correlations, and eventually  the unstable growth will shut-off.
When this happens, the state may be described by correlated domains with equal
probabibility for both phases inside the domains. The expectation value of
the field in this configuration will be zero, but inside each domain, the
field will acquire a value very close to the value in equilibrium at the
minimum of the {\it effective potential}. The size of the domain in this
picture will depend on the time during which correlations
had grown enough so that fluctuations start sampling beyond the spinodal point.

Since this physical picture may not be studied within perturbation theory,
we now introduce a {\it non-perturbative} method based on a
self-consistent Hartree approximation.

This approximation is implemented as follows:
in the initial Lagrangian write
\begin{equation}
\frac{\lambda}{4!}\Phi^4(\vec{r},t) =
\frac{\lambda}{4}\langle\Phi^2(\vec{r},t)\rangle
\Phi^2(\vec{r},t)+
\left(\frac{\lambda}{4!}\Phi^4(\vec{r},t)-
\frac{\lambda}{4}\langle\Phi^2(\vec{r},t)\rangle\Phi^2(\vec{
r},t)\right)
\label{hart}
\end{equation}
the first term is absorbed in a shift of the mass term
\[m^2(t) \rightarrow
m^2(t)+\frac{\lambda}{2}\langle\Phi^2(t)\rangle \]
(where we used spatial translational invariance). The second
term in
(\ref{hart}) is taken as an interaction with the term
$\langle\Phi^2(t)\rangle\Phi^2(\vec{r},t)$ as a mass counterterm.
The Hartree approximation consists of  requiring that the one loop
correction to the two point Green's functions must be cancelled by the mass
counterterm. This leads to the self consistent set of equations
\begin{equation}
\langle\Phi^2(t)\rangle  =  \int
\frac{d^3k}{(2\pi)^3}\left(-
iG_k^{<}(t,t)\right) =
\int \frac{d^3k}{(2\pi)^3} \frac{1}{2\omega_{<}(k)} \;
{\cal{U}}^+_k(t)
{\cal{U}}^-_k(t) \; \coth[\beta_i\omega_{<}(k)/2] \label{fi2}
\end{equation}
\begin{equation}
\left[\frac{d^2}{dt^2}+\vec{k}^2+m^2(t)+\frac{\lambda}{2}\langle
\Phi^2(t)\rangle\right]
{\cal{U}}^{\pm}_k=0 \label{hartree}
\end{equation}

Before proceeding any further, we must address the fact that
the composite
operator $\langle\Phi^2(\vec{r},t)\rangle$ needs one
subtraction and
multiplicative
renormalization. As usual the subtraction is absorbed in a
renormalization
of the bare mass, and the multiplicative renormalization into a
renormalization of the coupling constant. We must also point out that the
Hartree approximation is uncontrolled in this scalar theory; it becomes
equivalent to the large-N limit in theories in which the field is in the vector
representation of O(N) (see for example\cite{dolan}).

At this stage our justification for using this approximation
is based on the fact that it provides a non-perturbative framework to sum
an infinite series of Feynman diagrams of the cactus type\cite{dolan,chang}.

In principle one may improve on this approximation by using
the Hartree propagators in a loop expansion. The cactus-type diagrams will
still be cancelled by the counterterms (Hartree condition), but other
diagrams with loops (for example diagrams with multiparticle thresholds)
may be computed by using the Hartree propagators on the
lines. This approach will have the advantage that the Hartree propagators
will only be unstable for a finite time $ \tau \leq \tau_s $. It is not
presently
clear to these authors, however, what if any,  would be the expansion
parameter in this case.

It is clear that for $t<0$ there is a self-consistent
solution to the
Hartree equations with equation (\ref{fi2}) and
\begin{eqnarray}
\langle\Phi^2(t)\rangle       & = &  \langle\Phi^2(0^-)\rangle \nonumber \\
{\cal{U}}^{\pm}_k             & = &  \exp[\mp i \omega_{<}(k)t] \\
\omega^2_{<}(k)               & = &  \vec{k}^2+m^2_i+\frac{\lambda}{2}+
\langle\Phi^2(0^-)\rangle = \vec{k}^2+m^2_{i,R} \nonumber \\
\end{eqnarray}

where the composite operator has been absorbed in a renormalization of the
initial mass, which is now parametrized as
$m^2_{i,R}=\mu^2_R[(T^2_i/T^2_c)-1]$. For
$t>0$ we subtract the composite operator at $\tau=0$ absorbing the subtraction
into a renormalization of $m^2_f$ which we now parametrize as $m^2_{f,R}=
\mu^2_R[1-(T^2_f/T^2_c)]$. We should point out that this choice of
parametrization only represents a choice of the bare parameters, which can
always be chosen to satisfy this condition. The logarithmic multiplicative
divergence of the composite operator will be absorbed in a coupling constant
renormalization consistent with the Hartree approximation,
however, for the purpose of understanding the dynamics of growth of
instabilities associated with the long-wavelength fluctuations,
we will not need to specify this procedure. After this subtraction
procedure, the Hartree equations read
\begin{equation}
\langle\Phi^2(t)\rangle-\langle\Phi^2(0)\rangle  =
\int \frac{d^3k}{(2\pi)^3} \frac{1}{2\omega_{<}(k)} \;
[{\cal{U}}^+_k(t) {\cal{U}}^-_k(t)-1] \; \coth[\beta_i\omega_{<}(k)/2]
\label{subfi2}
\end{equation}
\begin{equation}
\left[\frac{d^2}{dt^2}+\vec{k}^2+m^2_R(t)+\frac{\lambda_R}{2}
\left(\langle\Phi^2(t)\rangle-\langle\Phi^2(0)\rangle\right)
\right]
{\cal{U}}^{\pm}_k(t)=0 \label{subhartree}
\end{equation}
\begin{equation}
m^2_R(t)= \mu^2_R \left[\frac{T^2_i}{T^2_c}-1\right]
\Theta(-t)
- \mu^2_R \left[1-\frac{T^2_f}{T^2_c}\right] \Theta(t)
\end{equation}
with $T_i > T_c$ and $T_f \ll T_c$.
With the self-consistent solution and boundary condition for
$t<0$
\begin{eqnarray}
\langle\Phi^2(t<0)\rangle-\langle\Phi^2(0)\rangle  & = & 0
\label{bcfi2} \\
{\cal{U}}^{\pm}_k(t<0)       & = & \exp[\mp i
\omega_{<}(k)t]
\label{bcmodes}\\
\omega_{<}(k)                & = & \sqrt{\vec{k}^2+m^2_{iR}}
\end{eqnarray}

This set of Hartree equations is extremely complicated
to be solved exactly.
However it has the correct physics in it. Consider the
equations for $t>0$,
at very early times, when (the renormalized) $\langle\Phi^2(t)\rangle-
\langle\Phi^2(0)\rangle \approx 0$
the mode functions are the same as in the zeroth order approximation,
and the unstable modes grow exponentially. By computing the expression
(\ref{subfi2}) self-consistently  with
these zero-order unstable modes, we see that the fluctuation
operator begins to grow exponentially.

As $(\langle\Phi^2(t)\rangle-\langle\Phi^2(0)\rangle)$ grows
larger,
its contribution to the Hartree equation tends to balance
the negative
mass term, thus weakening the unstabilities, so that only longer
wavelengths can become
unstable. Even for very weak coupling constants,
the exponentially
growing modes make the Hartree term in the equation of
motion for the mode
functions become large and compensate for the negative mass
term.
Thus when

\[\frac{\lambda_R}{2m^2_{f,R}}\left(\langle\Phi^2(t)\rangle-
\langle\Phi^2(0)\rangle\right)
\approx 1 \]
the instabilities
shut-off, this equality determines the spinodal time $\tau_s$.
The modes will still continue to grow further
after this point
because the time derivatives are fairly (exponentially)
large, but eventually
the growth will slow-down when fluctuations sample deep
inside the stable  region.

After the subtraction, and multiplicative renormalization
(absorbed in a
coupling constant renormalization), the composite operator
is finite. The
stable mode functions will make a {\it perturbative}
contribution to the
fluctuation which will be always bounded in time.  The most
important contribution will be that of the {\it unstable
modes}. These will grow
exponentially at early times and their effect will dominate
the dynamics of
growth and formation of correlated domains. The full set of
Hartree equations
is extremely difficult to solve, even numerically, so we
will restrict
ourselves to account {\it only} for the unstable modes. From
the above
discussion it should be clear that these are the only
relevant modes for the
dynamics of  formation and growth  of domains, whereas the
stable modes,  will
always contribute perturbatively for weak coupling after renormalization.

Introducing the dimensionless ratios (\ref{dimensionless}) in terms of
$m_{f,R}\; ; \; m_{i,R}$, (all momenta are now expressed in units of
$m_{f,R}$), dividing (\ref{subhartree}) by $m_{f,R}^2$,
using the high temperature
approximation $\coth[\beta_i\omega_{<}(k)/2] \approx 2T_i/\omega_{<}(k)$
for the unstable modes, and expressing the critical temperature as
$T^2_c=24 \mu_R/\lambda_R$, the set of Hartree equations
(\ref{subfi2},
\ref{subhartree}) become the following integro-differential equation for
the mode functions for $t>0$
\begin{equation}
\left[\frac{d^2}{d\tau^2}+q^2-1+g\int^1_0 dp
\left\{\frac{p^2}{p^2+L^2_R}
[{\cal{U}}^+_p(t){\cal{U}}^-_p(t)-1]\right\}
\right]{\cal{U}}^{\pm}_q(t)=0
\label{finalhartree}
\end{equation}
with
\begin{eqnarray}
{\cal{U}}^{\pm}_q(t<0) & = & \exp[\mp i \omega_{<}(q)t]
\label{bounconhart} \\
\omega_{<}(q)          & = & \sqrt{q^2+L^2_R} \label{frequ}
\\
L^2_R                  & = & \frac{m^2_{i,R}}{m^2_{f,R}} =
\frac{T_i^2-T_c^2}{T^2_c-T_f^2}
\\
g                      & = & \frac{\sqrt{24\lambda_R}}{4\pi^2}
\frac{T_i}{[T^2_c-T^2_f]^{\frac{1}{2}}}
\label{effectivecoupling}
\end{eqnarray}
The effective coupling (\ref{effectivecoupling}) reflects
the enhancement of
quantum fluctuations by high temperature effects; for
$T_f/T_c \approx 0$,
and for couplings as weak as $\lambda_R \approx 10^{-12}$,
$g \approx 10^{-7} (T_i/T_c)$.

The equations (\ref{finalhartree}) may now be integrated
numerically for the
mode functions; once we find these, we can then compute the
contribution of the unstable modes
 to the subtracted
correlation function equivalent to  (\ref{integral})
\begin{eqnarray}
{\cal{D}}^{(HF)}(x,\tau)    & = &
\frac{\lambda_R}{6 m_{f,R}^2} \left[\langle \Phi(\vec{r},t)
\Phi(\vec{0},t)\rangle-
\langle\Phi(\vec{r},0)\Phi(\vec{0},0)\rangle\right]
\label{hartreecorr1} \\
3{\cal{D}}^{(HF)}(x,\tau)   & = & g\int_0^1dp
\frac{p^2}{p^2+L^2_R}
\frac{\sin(px)}{px}\left[{\cal{U}}^+_p(t){\cal{U}}^
-_p(t)-1\right]
\label{hartreecorr2}
\end{eqnarray}
In figure (4) we show

\[ \frac{\lambda_R}{2m_{f,R}^2}(\langle\Phi^2(\tau)\rangle -
\langle\Phi^2(0)\rangle)=
3({\cal{D}}^{HF}(0,\tau)- {\cal{D}}^{HF}(0,0)) \] (solid line) and
also for comparison, its zeroth-order counterpart
$3({\cal{D}}^{(0)}(0,\tau)-{\cal{D}}^{(0)}(0,0))$ (dashed line)
for $\lambda_R = 10^{-12}\; , \; T_i/T_c=2$.
(Again, this value of the initial temperature does not have any
particular physical significance and was chosen as a
representative).  We clearly see what we expected;
 whereas the zeroth order correlation grows indefinitely,
the Hartree
correlation function is bounded in time and oscillatory. At
$\tau \approx
10.52$ ,  $3({\cal{D}}^{(HF)}(0,\tau)-{\cal{D}}^{(HF)}(0,\tau))= 1$,
fluctuations are sampling field configurations near the
classical spinodal, fluctuations continue to grow  because the
derivatives are still fairly large. However, after this time, the
modes  begin  to probe the stable region in which there is no
exponential growth. At this point
$\frac{\lambda_R}{2m_{f,R}^2}(\langle\Phi^2(\tau)\rangle-\Phi^2(0)\rangle)$,
becomes small again because of the small coupling $g \approx
10^{-7}$, and the correction term diminishes.  When it becomes
smaller than one, the instabilities set in again, modes
begin to grow and the process repeats.
This gives rise to an oscillatory behavior around
$\frac{\lambda_R}{2m^2_{f,R}}(\langle\Phi^2(\tau)\rangle-
\Phi^2(0)\rangle=1$ as shown in figure (5).

In figures (6.a-d), we show the structure factors as a
function of
$x$ for $\tau = 6, \; 8, \; 10, \; 12$, both for zero-order
(tree
level) ${\cal{D}}^{(0)}$ (dashed lines) and Hartree
${\cal{D}}^{(HF)}$ (solid lines).
These correlation functions clearly show that correlations
grow in amplitude
 and that the size of the region in which
the fields are
correlated increases with time. Clearly this region may be
interpreted as
a ``domain'', inside which the fields have strong
correlations, and outside
which the fields are uncorrelated.

We see that up to the spinodal time $\tau_s \approx 10.52$
at which
 $\frac{\lambda_R}
{2m_{f,R}}(\langle\Phi^2(\tau_s)\rangle-
\Phi^2(0)\rangle)=1$, the zeroth order
correlation
$3{\cal{D}}^{(0)}(0,\tau)$ is very close to the Hartree
result. In fact at
$\tau_s$, the difference is less than $15\%$.  In particular for
these values of the coupling and initial temperature, the zeroth
order correlation function leads to $\tau_s \approx 10.15$, and we may
use the zeroth order correlations to provide an analytic estimate for
$\tau_s$, as well as  the form of the correlation functions and the
size of the  domains.

The fact that the zeroth-order
correlation remains very close to the Hartree-corrected
correlations up to
times comparable to the spinodal is clearly a consequence of
the very small coupling.

To illustrate this fact, we show in figures (7,8) the same
correlation functions
but for $\lambda = 0.01 , \; T_i/T_c =2$. Clearly the
stronger coupling makes
the growth of domains much faster and the departure from
tree-level
correlations more dramatic. Thus, it becomes clear that for
strong couplings,
domains will form very rapidly and only grow to sizes of the
order of the
zero temperature correlation length. The phase transition
will occur very
rapidly, and clearly our initial assumption of a rapid
supercooling will be unjustified.
 This situation for strong couplings, of domains
forming very
rapidly to sizes of the order of the zero temperature
correlation length,
is the picture presented by Mazenko and collaborators\cite{mazenko}.
However, for very
weak couplings (consistent with the bounds from density
fluctuations), our
results indicate that the phase transition will proceed very
slowly, domains
will grow for a long time and become fairly large,
with a typical size several times the zero
temperature correlation length. In a sense, this is a self
consistent check
of our initial assumptions on a rapid supercooling in the
case of weak couplings.

Thus, as we argued above, for very weak coupling,
we may use the tree level result to give an
approximate bound to the correlation functions up to times
close to the
spinodal time using the result given by equation (\ref{strucfactau}),
for $T_f \approx 0$.
Thus, we conclude that for large times, and very weakly
coupled theories ($\lambda_R \leq 10^{-12}$) and for initial temperatures
of the order of the critical temperature,
the size of the domains $\xi_D(t)$ will grow typically in time
as

\begin{equation}
 \xi_D(t)       \approx  (8\sqrt{2})^{\frac{1}{2}}
 \xi(0)\sqrt{\frac{t}{\xi(0)}}
\label{sizedomain}
 \end{equation}

with $\xi(0)$ the zero temperature correlation length. The
maximum size of
a domain is approximately determined by the time at which
fluctuations begin
probing the stable region, this is the spinodal time $\tau_s$ and the
maximum size of the domains is approximately $\xi_D(\tau_s)$.

An estimate for the spinodal time, is obtained from equation
(\ref{strucfactau}) by the condition $3{\cal{D}}(\tau_s)=1$,
then for weakly
coupled theories and $T_f \approx 0$, we obtain
\begin{equation}
\tau_s = -\ln\left[
\left(\frac{3\lambda}{4\pi^3}\right)^
{\frac{1}{2}}\frac{(\frac{T_i}{2 T_c})^3}{
\frac{T^2_i}{T^2_c}-1}\right)
\end{equation}

It is remarkable that the domain size scales as $\xi_D(t)
\approx t^{\frac{1}{2}}$ just like in classical theories
 of spinodal decomposition,
when the order parameter {\it is not conserved}, as is
the case in the scalar relativistic field theory under consideration,
but certainly for completely different reasons\cite{boyspino}. At the tree
level, we can identify this scaling behavior as arising from the
relativistic dispersion
relation, and second order time derivatives in the equations of motion, a
situation very different from the classical description of
the Allen-Cahn-Lifshitz\cite{langer,guntonmiguel} theory of spinodal
decomposition based on a time-dependent Landau-Ginzburg model.

When the scalar coupling is strong, the phase transition proceeds rapidly,
and domains will not have time to grow substantially. Their sizes will be
of the order of the  zero temperature correlation length. This will become
important in a later section when we study the dynamics in a phenomenologically
relevant strongly coupled theory.

{\bf Beyond Hartree:}

The Hartree approximation, keeping only the unstable modes
in the self-
consistent equation, clearly cannot be accurate for times
beyond the
spinodal time. When the oscillations in the Hartree solution
begin, the field
fluctuations are probing the stable region. This should
correspond to the
onset of the  ``reheating'' epoch, in which dissipative
effects become
important for processes of particle and entropy production.
Clearly
the Hartree approximation ignores all dissipative processes,
as may be
understood from the fact that this approximation sums the
cactus type
diagrams for which there are no multiparticle thresholds.
Furthermore, in
this region, the contribution of the stable modes to the
Hartree equation
becomes important for the subsequent evolution beyond the
spinodal point and
clearly will contribute to the ``reheating'' process.
A possible approach to incorporate the contribution of the
stable modes may be that explored by Avan and de
 Vega\cite{avan} in terms of the effective action for the
 composite operator.

Thus, although the Hartree approximation may give a fairly
accurate picture
of the process of domain formation and growth, one must go
beyond this
approximation at times later than the spinodal time, to
incorporate
dissipative  effects and to study the ``reheating'' period.
Clearly, one must also attempt to study the possibility of
``percolation of domains''. Furthermore, the Hartree
approximation is essentially a Gaussian approximation, as
the wave-functional (or in this case the functional density
matrix) is Gaussian with kernels that are obtained self-
consistently. The wave-functional must include non-gaussian
correlations that will account for the corrections to the
Hartree approximation and will be important to obtain the long time
behavior for $\tau > \tau_s$.

\section{\bf Non-Equilibrium Dynamics in FRW Cosmologies}

In this section we provide an alternative and equally powerful method to
obtain the evolution equations by evolving in time an initially prepared
density matrix. This method is fundamentally equivalent to the one described
in the previous section, but is restricted to gaussian density matrices.
This restriction, however still offers the possibility of studying both
perturbative and non-perturbative approximations and we include it here
as one more tool in non-equilibrium field theory. Furthermore this method
permits the study of effects of particle production via parametric
amplification in a more natural way.

What we address in this section is the effective evolution equations for the
order parameter and quantum fluctuations in
homogeneous and isotropic FRW expanding cosmologies.

By and large, the
various models of inflation make the assumption that the dynamics of the
spatial zero mode of the (so-called) inflaton field is governed by some
approximation to the effective potential which incorporates the effects of
quantum fluctuations of the field. Thus the equation of motion is usually of
the form:

\begin{equation}
\ddot{\phi} + 3\frac{\dot{a}}{a}\dot{\phi} + V'_{\rm eff}(\phi) = 0
\end{equation}

The problem here is that the effective potential is really only suited for
analyzing {\em static} situations; it is the effective action evaluated for a
field configuration that is constant in time. Thus, it
is inconsistent to use the effective potential in a {\em dynamical}
situation. Notice that such inconsistency appears for {\bf any}
inflationary scenario (old, new, chaotic, ...).

Our approach is to use the functional Schr\"{o}dinger formulation, wherein we
specify the initial wavefunctional $\Psi[\Phi(\vec{.}); t]$
(or more generally a
density matrix $\rho[\Phi(\vec{.}), \tilde{\Phi}(\vec{.}); t]$), and then use
the Schr\"{o}dinger equation to evolve this state in time. We can then use
this state to compute all of the expectation values required in the
construction of the effective equations of motion for the order parameter of
the theory, as well as that for the fluctuations. The Schr\"{o}dinger approach
has already been used in the literature at zero temperature\cite{hill} and to
study non-equilibrium aspects of field theories\cite{eboli,cooper}.

We start by setting up the {Schr\"{o}dinger} formalism for spatially flat FRW
cosmologies.
Consider a scalar field in such a cosmology where the metric is:
\begin{equation}
ds^2 = dt^2-a^2(t)d\vec{x}^2
\end{equation}
 The action and Lagrangian density are given by
\begin{eqnarray}
S         & =  & \int d^4x {\cal{L}} \label{action} \\
{\cal{L}} & =  & a^3(t)\left[\frac{1}{2}\dot{\Phi}^2(\vec{x},t)-\frac{1}{2}
\frac{(\vec{\nabla}\Phi(\vec{x},t))^2}{a(t)^2}-V(\Phi(\vec{x},t))\right]
 \label{lagrangian} \\
V(\Phi)   & =  & \frac{1}{2}[m^2+ \xi {\cal{R}}] \Phi^2(\vec{x},t)+
\frac{\lambda}{4!}\Phi^4(\vec{x},t) \label{potential} \\
{\cal{R}}    & =  & 6\left(\frac{\ddot{a}}{a}+\frac{\dot{a}^2}{a^2}\right)
\end{eqnarray}
with ${\cal{R}}$ the Ricci scalar.
The canonical momentum conjugate to $\Phi$ is
\begin{equation}
\Pi(\vec{x},t) = a^3(t)\dot{\Phi}(\vec{x},t) \label{canonicalmomentum}
\end{equation}
and the Hamiltonian becomes
\begin{equation}
H(t) = \int d^3x \left\{ \frac{\Pi^2}{2a^3(t)}+
\frac{a(t)}{2}(\vec{\nabla}\Phi)^2+
a^3(t) V(\Phi) \right\} \label{hamiltonian}
\end{equation}
In the Schr\"{o}dinger representation (at an arbitrary fixed time
$t_o$), the canonical momentum is represented
as
\[ \Pi(\vec{x}) = -i\hbar \frac{\delta}{\delta \Phi(\vec{x})} \]
Wave functionals obey the time dependent functional Schr\"{o}dinger equation
\begin{equation}
i\hbar \frac{\partial \Psi[\Phi,t]}{\partial t} = H \Psi[\Phi,t]
 \label{schroedinger}
\end{equation}

Since we shall eventually consider a ``thermal ensemble'' it is convenient to
work with a functional density matrix $\hat{\rho}$ with matrix elements in the
Schr\"{o}dinger representation
$\rho[\Phi(\vec{.}), \tilde{\Phi}(\vec{.});t]$. We will
{\it assume} that the
density matrix obeys the functional Liouville equation
\begin{equation}
i\hbar \frac{\partial \hat{\rho}}{\partial t} = \left[H(t),\hat{\rho}\right]
\label{liouville}
\end{equation}
whose formal solution is
\[ \hat{\rho}(t) = U(t,t_o) \hat{\rho}(t_o) U^{-1}(t,t_o)\] where
$U(t,t_o)$ is the time evolution operator, and $\hat{\rho}(t_o)$ the
density matrix at the arbitrary initial time $t_o$.

The diagonal density matrix elements $\rho[\Phi,\Phi;t]$ are interpreted
as a probability density in functional space.
Since we are considering an homogeneous and isotropic background, the
functional density matrix may be assumed to be translationally invariant.
 Normalizing the density matrix such that $Tr\hat{\rho}=1$, the
order parameter is defined as
\begin{equation}
\phi(t) = \frac{1}{\Omega}\int d^3x \langle \Phi(\vec{x},t) \rangle =
\frac{1}{\Omega}\int d^3x  Tr\left[\hat{\rho}(t)\Phi(\vec{x})\right] =
\frac{1}{\Omega}\int d^3x
Tr \left[\hat{\rho}(t_o)U^{-1}(t,t_o)\Phi(\vec{x},t_o)U(t,t_o)\right]
\label{orderparameter}
\end{equation}
where $\Omega$ is the comoving volume, and the scale factors cancel
between the numerator (in the integral) and the denominator. Note that we have
used the fact that the field operator does not evolve in time in this picture.
The evolution equations for the order parameter are the following
\begin{eqnarray}
\frac{d \phi(t)}{dt} & = & \frac{1}{a^3(t)\Omega}\int d^3x
  \langle \Pi(\vec{x},t) \rangle
 =\frac{1}{a^3(t)\Omega}\int d^3x  Tr\hat{\rho}(t)\Pi(\vec{x}) = \frac{\pi(t)}
{a^3(t)} \label{fidot} \\
\frac{d \pi(t)}{dt}     & = & -\frac{1}{\Omega}\int d^3x a^3(t) \langle
 \frac{\delta V(\Phi)}{\delta \Phi(\vec{x})} \rangle \label{pidot}
\end{eqnarray}
It is now convenient to write the field in the {Schr\"{o}dinger} picture as
\begin{eqnarray}
\Phi(\vec{x})   & = & \phi(t)+\eta(\vec{x},t) \label{split} \\
\langle \eta(\vec{x},t) \rangle
     & = & 0 \label{doteta}
\end{eqnarray}

Expanding the right hand side of (\ref{pidot}) we find the effective
equation of motion for the order parameter:
\begin{equation}
\frac{d^2 \phi(t)}{dt^2}+3 \frac{\dot{a}(t)}{a(t)}
\frac{d \phi(t)}{dt}+V'(\phi(t))+\frac{V'''(\phi(t))}{2 \Omega}\int d^3x
\langle \eta^2(\vec{x},t)\rangle+\cdots
=0 \label{effequation}
\end{equation}
where primes stand for derivatives with respect to $\phi$.
To leading order in the loop expansion we need that
$\langle \eta^2(\vec{x},t)\rangle = {\cal{O}}(\hbar)$. This will be guaranteed
to this order if the density matrix is assumed to be Gaussian with
a covariance (width) ${\cal{O}}(1/\hbar)$.
If (\ref{split}) is introduced in the Hamiltonian, we arrive at:
\begin{equation}
H(t) = \int d^3x \left\{-\frac{\hbar^2}{2 a^3(t)}\frac{\delta^2}{\delta
\eta^2}+\frac{a(t)}{2}\left(\vec{\nabla}\eta\right)^2+a^3(t)
\left(V(\phi)+V'(\phi)\eta+\frac{1}{2}V''(\phi)\eta^2+\cdots \right)
\right\} \label{quadham}
\end{equation}

Keeping only the terms quadratic in $\eta$ in (\ref{quadham})
gives the first order term in the loop expansion.

It is convenient to introduce the discrete Fourier transform of the
fields in the comoving frame as
\begin{equation}
\eta(\vec{x},t) = \frac{1}{\sqrt{\Omega}}\sum_{\vec{k}} \eta_{\vec{k}}(t)
e^{-i\vec{k}\cdot\vec{x}}
\label{fourier1}
\end{equation}
In this representation, the quadratic approximation to the Hamiltonian
(\ref{quadham}) becomes the Hamiltonian of a collection of independent harmonic
oscillators for each mode $\vec{k}$
\begin{eqnarray}
H_q & = & \Omega a^3(t) V(\phi(t))+\nonumber \\
 &   & \frac{1}{2} \sum_{\vec{k}}
\left\{ -\frac{\hbar^2}{a^3(t)}
 \frac{\delta^2}{\delta \eta_{\vec{k}}\delta\eta_{-\vec{k}}}+2a^3(t)
 V'_{\vec{k}}(\phi(t))\eta_{-\vec{k}}+
\omega^2_k(t)\eta_{\vec{k}}\eta_{-\vec{k}}\right\} \label{hamodes} \\
V'_{\vec{k}}(\phi(t))
 & = & V'(\phi(t))\sqrt{\Omega}\delta_{\vec{k},0} \nonumber \\
\omega^2_k(t)
 & = & a(t)\vec{k}^2+a^3(t)V''(\phi(t)) \label{timedepfreq}
\end{eqnarray}

We propose the following Gaussian ansatz for the functional density
matrix elements in the {Schr\"{o}dinger} representation
\begin{eqnarray}
\rho[\Phi,\tilde{\Phi},t] & = & \prod_{\vec{k}} {\cal{N}}_k(t) \exp\left\{
- \left[\frac{A_k(t)}{2\hbar}\eta_k(t)\eta_{-k}(t)+
\frac{A^*_k(t)}{2\hbar}\tilde{\eta}_k(t)\tilde{\eta}_{-k}(t)+
\frac{B_k(t)}{\hbar}\eta_k(t)\tilde{\eta}_{-k}(t)\right] \right. \nonumber \\
       &   & \left. +\frac{i}{\hbar}\pi_k(t)\left(\eta_{-k}(t)-
\tilde{\eta}_{-k}(t)\right) \right\} \label{densitymatrix} \\
\eta_k(t)          & = & \Phi_k-\phi(t)\sqrt{\Omega}\delta_{\vec{k},0}
\label{etaofkt} \\
\tilde{\eta}_k(t)       & = &
{\tilde{\Phi}}_k-\phi(t)\sqrt{\Omega}\delta_{\vec{k},0}
\label{etaprimeofkt}
\end{eqnarray}
where $\phi(t) = \langle \Phi(\vec{x}) \rangle$ and $\pi_k(t)$ is the Fourier
transform of $\langle \Pi(\vec{x}) \rangle$. This form of the density matrix
is dictated by the hermiticity condition $\rho^{\dagger}[\Phi,\tilde{\Phi},t] =
\rho^*[\tilde{\Phi},\Phi,t]$; as a result of this, $B_k(t)$ is real.
The kernel $B_k(t)$ determines the amount of ``mixing'' in the
density matrix, since if $B_k=0$, the density matrix corresponds to a pure
state because it is a wave functional times its complex conjugate.

In order to solve for the time evolution of the density matrix
(\ref{liouville}) we need to specify the density matrix at some initial
time $t_o$. It is at this point that we have to {\it assume} some physically
motivated initial condition. We believe that this is a subtle point that
has not received proper consideration in the literature. A system in
thermal equilibrium has time-independent ensemble averages (as the evolution
Hamiltonian commutes with the density matrix) and there is no memory of any
initial state. However, in a time dependent background, the density matrix
will evolve in time, departing from the equilibrium state and
correlation functions or expectation values may depend on
details of the initial state.

We will {\it assume} that at early times
the initial density matrix is {\em thermal} for the modes that
diagonalize the
Hamiltonian at $t_o$ (we call these the {\em adiabatic} modes). The effective
temperature for these modes is $k_BT_o = 1/\beta_o$. It is only in this
initial state that the notion of ``temperature'' is meaningful. As the
system departs from equilibrium one cannot define a thermodynamic temperature.
Thus in this case the ``temperature'' refers to the temperature defined in the
initial state.

The initial values of the
order parameter and average canonical momentum are $\phi(t_o) =\phi_o$ and
$\pi(t_o)=\pi_o$ respectively. Defining the adiabatic frequencies as:
\begin{equation}
 W^2_k(t_o) = \frac{\omega^2_k (t_o)}{a^3(t_o)}=
\frac{\vec{k}^2}{a^2(t_o)}+V''(\phi_{cl}(t_o)) \label{adiabfreq}
\end{equation}
we find that the initial values of the time dependent
parameters in the density matrix (\ref{densitymatrix}) are
\begin{eqnarray}
A_k(t_o) & = & A^*_k(t_o) =
 W_k(t_o)a^3(t_o)
\coth\left[\beta_o\hbar W_k(t_o)
\right] \label{Ato} \\
B_k(t_o) & = & - \frac{W_k(t_o)a^3(t_o)}
{\sinh\left[\beta_o\hbar W_k(t_o)\right]}
\label{Bto} \\
{\cal{N}}_k(t_o)
   & = &  \left[\frac{W_k(t_o)a^3(t_o)}{\pi\hbar}\tanh\left[
\frac{\beta_o\hbar W_k(t_o)}{2}\right]\right]^{\frac{1}{2}}
 \label{normo} \\
\phi(t_o)& = & \phi_o \; \; ; \; \; \pi(t_o)=\pi_o
\label{initialfielmom}
\end{eqnarray}
The initial density matrix is normalized such that
$Tr\rho(t_o)=1$. Since time evolution is unitary such a normalization
will be constant in time. For $T_o = 0$ the density matrix describes a
pure state since $B_k = 0$.

As an example, consider the case of de Sitter space. The scale factor is
given by $a(t) =a_o e^{Ht}$ and for $T_o \rightarrow 0$, $t_o \rightarrow
-\infty$ we recognize the
ground state wave-functional for the Bunch-Davies vacuum\cite{hill,birrell}.
For $T_o \neq 0$ this
initial density matrix corresponds to a thermal ensemble of
Bunch-Davies modes.
Certainly this choice is somewhat arbitrary but it physically describes the
situation in which at very early times the adiabatic modes are in {\it local
thermodynamic equilibrium}. Whether or not this situation actually obtains for
a given system has to be checked explicitly. In the cosmological setting,
the nature of the initial condition will necesarily have to result from a
deeper understanding of the relationship between particle physics, gravitation
and statistical mechanics at very large energy scales.

Although we will continue henceforth to use this thermal initial state, it
should be emphasized that our formalism is quite general and can be applied to
{\em any initial} state.

In the {Schr\"{o}dinger} picture, the Liouville equation (\ref{liouville})
becomes
\begin{eqnarray}
& & i\hbar \frac{\partial \rho[\Phi,\tilde{\Phi},t]}{\partial t} =
\sum_k\left\{
-\frac{\hbar^2}{2a^3(t)}\left(
\frac{\delta^2}{\delta \eta_k \delta \eta_{-k}}-
\frac{\delta^2}{\delta \tilde{\eta}_k \delta \tilde{\eta}_{-k}}\right) \right.
 \nonumber \\
& & \left. +a^3(t){V'}_{-k}(\phi(t))\left(\eta_{k}-\tilde{\eta}_k\right)+
\frac{1}{2}\omega^2_k(t)\left(\eta_{\vec{k}}\eta_{-\vec{k}}
-\tilde{\eta}_k\tilde{\eta}_{-k}\right)
\right\} \rho[\Phi,\tilde{\Phi},t] \label{liouvischroed}
\end{eqnarray}
Since the modes do not mix in this approximation to the Hamiltonian, the
equations for the kernels in the density matrix are obtained by comparing the
powers of $\eta$ on both sides of the above equation. We obtain the following
equations for the coefficients:
\begin{eqnarray}
i\frac{\dot{{\cal{N}}}_k}{{\cal{N}}_k} & = & \frac{1}{2a^3(t)}(A_k-A^*_k)
\label{normeq} \\
i\dot{A}_k                     & = &
\left[ \frac{A^2_k-B^2_k}{a(t)^3}-\omega^2_k(t)\right] \label{Aeq} \\
i \dot{B_k}                    & = & \frac{B_k}{a^3(t)}(A_k-A^*_k)
\label{Beq} \\
-\dot{\pi}_k                   & = & V'(\phi(t))a^3(t) \sqrt{\Omega}
\delta_{\vec{k},0} \label{pieq} \\
\dot{\phi}                     & = & \frac{\pi}{a^3(t)}
\label{fieq}
\end{eqnarray}
The last two equations are identified with the {\it classical}
 equations of motion for the order parameter (\ref{effequation}).
The equation for $B_k(t)$ reflects the fact that a pure state
$B_k=0$ remains pure under time evolution.

Writing $A_k$ in terms of its real and imaginary components
$A_k(t) = A_{Rk}(t)+i A_{Ik}(t)$ (and because $B_k$ is real) we
find that
\begin{equation}
\frac{B_k(t)}{A_{Rk}(t)} = \frac{B_k(t_o)}{A_{Rk}(t_o)}
\label{invar}
\end{equation}
and that  the time evolution is unitary (as it should be), that is
\begin{equation}
\frac{{\cal{N}}_k(t)}{\sqrt{\left(A_{Rk}(t)+B_k(t)\right)}} = \mbox{constant}
\label{unitarity}
\end{equation}

The initial conditions (\ref{Ato},\ref{Bto}) and the
invariance of the ratio (\ref{invar}) suggest that
the solution for the real part of $A$ and for $B$ may be obtained by
introducing a complex function ${\cal{A}}_k(t)={\cal{A}}_{Rk}(t)+i
{\cal{A}}_{Ik}(t)$
\begin{eqnarray}
{\cal{A}}_{Rk}(t)           & = & A_{Rk}(t)
\tanh\left[\beta_o\hbar W_k(t_o)\right] =
- B_k(t)\sinh\left[\beta_o\hbar W_k(t_o)\right]  \label{Aoftime} \\
{\cal{A}}_{Rk}(t_o)     & = & W_k(t_o)~a^3(t_o) \label{ARinitial} \\
{\cal{A}}_{Ik}(t)               & = & A_{Ik}(t) \label{imagpart}
\end{eqnarray}
In this form, the real and imaginary parts of ${\cal{A}}$ satisfy the
equations
\begin{eqnarray}
\dot{{\cal{A}}}_{Rk}(t)  & = & \frac{2}{a^3(t)}{\cal{A}}_{Rk}(t)
{\cal{A}}_{Ik}(t) \label{realeq} \\
-\dot{{\cal{A}}}_{Ik}(t) & = & \frac{1}{a^3(t)}\left[{\cal{A}}^2_{Rk}(t)-
{\cal{A}}^2_{Ik}(t) -\omega^2_k(t)a^3(t)\right] \label{imageq}
\end{eqnarray}
These two equations may be combined in one complex equation for the
combination
 ${\cal{A}}_k(t) = {\cal{A}}_{Rk}(t)+
i{\cal{A}}_{Ik}(t)$ that obeys the Riccati-type equation
\begin{equation}
i\dot{\cal{A}}_k(t) = \frac{1}{a^3(t)}\left[{{\cal{A}}_k}^2(t)-
\omega^2_k(t)a^3(t) \right] \label{riccati}
\end{equation}
with the initial conditions:
\begin{eqnarray}
{\cal{A}}_{Rk}(t_o) & = &  W_k(t_o)a^3(t_o) \label{Are} \\
{\cal{A}}_{Ik}(t_o) & = & 0 \label{Aim}
\end{eqnarray}

The Riccati equation (\ref{riccati}) becomes a more amenable differential
equation by the change of variables

\begin{equation}
{\cal{A}}_k(t) = -ia^3(t)\frac{\dot{\varphi}_k(t)}{\varphi_k(t)}
\label{changeofvar}
\end{equation}

The solution to the Ricatti equation with the above initial conditions
is detailed in  appendix A. We find that it is convenient to introduce
two real mode functions (for each wavevector $k$) and write

\begin{equation}
\varphi_k(t) =
\frac{{\cal{U}}_{k1}(t)+i{\cal{U}}_{k2}(t)}{\sqrt{a^3(t)W_k(t_o)}}.
 \label{chanvar}
\end{equation}
These mode functions obey a {Schr\"{o}dinger}-like equation. The initial
conditions on $\varphi_k(t_o)$ are (see appendix A)

\begin{eqnarray}
\varphi_k(t_o)               & = & \frac{1}{\sqrt{a^3(t_o)W_k(t_o)}}
 \nonumber \\
\dot{\varphi}_k(t)\mid_{t_o} & = &
 i\sqrt{\frac{W_k(t_o)}{a^3(t_o)}}
\nonumber
\end{eqnarray}

 ${\cal{A}}_{Rk}(t)$ is given by equation
 (\ref{arealfin}) in  appendix A, so that:

\begin{eqnarray}
A_{Rk}(t) & = &
 \frac{1}{\mid \varphi_k(t) \mid^2}
\coth\left[\beta_o\hbar W_k(t_o)
\right] \label{Aoftimefin} \\
B_k(t)    & = & -
\frac{1}{\mid \varphi_k(t) \mid^2}
\left[\frac{1}{\sinh\left[\beta_o\hbar W_k(t_o)\right]}\right]
\label{Boftimefin}
\end{eqnarray}

The equal time two-point function for the fluctuation becomes
\begin{equation}
\langle \eta_{k}(t)\eta_{-k}(t) \rangle = \frac{\hbar}{2\left(
A_{Rk}(t)+B_{k}(t)\right)} =
\frac{\hbar}{2}{\mid \varphi_k(t) \mid^2}
\coth\left[\beta_o\hbar W_k(t_o)/2
\right] \label{twopointfunc}
\end{equation}
The one-loop equation of motion for the order parameter thus becomes

\begin{equation}
\ddot{\phi}+3 \frac{\dot{a}}{a}\dot{\phi}+V'(\phi)+V'''(\phi)
\frac{\hbar}{2}\int \frac{d^3k}{(2\pi)^3}
\frac{\mid \varphi_k(t) \mid^2}{2}
\coth\left[\beta_o\hbar W_k(t_o)/2
\right] = 0 \label{finaleqofmot}
\end{equation}
with the function $\varphi_k(t)$ defined  in  appendix A by   (\ref{calU},
\ref{diffeqU}) in which, to this order in $\hbar$, only the {\it classical
solution}  $\phi_{cl}(t)$ enters. A consistent numerical solution of these
equations to ${\cal{O}}(\hbar)$ would involve splitting
$\phi(t)=\phi_{cl}(t)+\hbar \phi_{(1)}(t)$ and keeping only the
${\cal{O}}(\hbar)$ terms in the evolution equation. This will result in two
simultaneous equations, one for the classical evolution of the order
parameter and another for $\phi_{(1)}(t)$.

This equation of motion is clearly {\em very} different from the one obtained
by using the effective potential. It may be easily seen (by writing the
effective action as the classical action plus the logarithm of the determinant
of the quadratic fluctuation operator) that this is the equation of motion
obtained by the variation of the one-loop effective action.

The {\it static} effective potential is clearly not the appropriate quantity to
use to describe scalar field dynamics in an expanding universe. Although there
may be some time regime in which the time evolution is slow and fluctuations
rather small, this will certainly {\it not} be the case at the onset of a phase
transition. As the phase transition takes place, fluctuations become dominant
and grow in time signaling the onset of long range
correlations\cite{boyveg,boydom}.

\subsection{\bf Hartree equations}

Motivated by our  studies in Minkowski space reported in the
previous section, in which
showed that the growth of correlation and enhancement of fluctuations during
a phase transition may not be described perturbatively, we now proceed to
obtaining the equations of motion in a Hartree approximation.
As argued in the previous section, this
approximation is non-perturbative in the sense that it sums up infinitely
many diagrams of the cactus-type\cite{dolan,chang}.

Vilenkin\cite{vilenkin} has previously studied a simplified version of the
Hartree approximation.

The Hartree self-consistent approximation is implemented as follows. We
decompose the field as in (\ref{split}), using a potential as in
(\ref{potential}). We find that the Hamiltonian becomes
\begin{eqnarray}
H & = &  \int d^3x \left\{-\frac{\hbar^2}{2 a^3(t)}\frac{\delta^2}
{\delta\eta^2}+\frac{a(t)}{2}\left(\vec{\nabla}\eta\right)^2+a^3(t)
\left(V(\phi)+V'(\phi)\eta+\frac{1}{2!}V''(\phi)\eta^2 \right. \right.
\nonumber \\
  & + &
\left. \left. \frac{1}{3!} \lambda \phi\eta^3+
\frac{1}{4!} \lambda \eta^4 \right)\right\} \label{fullham}
\end{eqnarray}

The Hartree approximation is obtained by assuming the factorization
\begin{eqnarray}
\eta^3 (\vec{x},t) &  \rightarrow & 3 \langle \eta^2(\vec{x},t) \rangle
\eta(\vec{x},t) \label{eta3} \\
\eta^4 (\vec{x},t) &  \rightarrow & 6 \langle \eta^2(\vec{x},t) \rangle
\eta^2(\vec{x},t) -3
\langle \eta^2(\vec{x},t) \rangle^2 \label{eta4}
\end{eqnarray}
where $\langle \cdots \rangle$ is the average using the time evolved density
matrix. This average will be determined self-consistently (see below).
Translational invariance shows that $\langle \eta^2(\vec{x},t) \rangle$ can
only be a function of time. This approximation makes the Hamiltonian quadratic
at the expense of a self-consistent condition. In the time independent
(Minkowski) case this approximation sums up all the ``daisy'' (or ``cactus'')
diagrams and leads to the self-consistent gap equation\cite{dolan}. In this
approximation the Hamiltonian becomes
\begin{eqnarray}
& & H = \Omega a^3(t) {\cal{V}}(\phi) +  \nonumber \\
& &  \int d^3x \left\{-\frac{\hbar^2}{2 a^3(t)}\frac{\delta^2}{\delta
\eta^2}+ \frac{a(t)}{2}\left(\vec{\nabla}\eta\right)^2
+a^3(t)\left({\cal{V}}^{(1)}(\phi)\eta+\frac{1}{2}
{\cal{V}}^{(2)}(\phi)\eta^2 \right)
\right\} \label{hartham}\\
& & {\cal{V}}(\phi) = V(\phi)- \frac{1}{8}\lambda \langle \eta^2 \rangle^2
\label{newV} \\
& & {\cal{V}}^{(1)}(\phi) = V'(\phi)+\frac{1}{2}\lambda\phi
\langle \eta^2 \rangle \label{newVprime} \\
& & {\cal{V}}^{(2)}(\phi) = V''(\phi)+ \frac{1}{2}
\lambda  \langle \eta^2 \rangle \label{newVdoubprim}
\end{eqnarray}

We can now introduce the Fourier transform of the field as in
(\ref{fourier1}).The Hamiltonian will have the same form as (\ref{hamodes}) but
the time dependent frequencies (\ref{timedepfreq}) and linear term in $\eta$
become
\begin{eqnarray}
\omega_k^2(t) & = & a(t) \vec{k}^2+a^3(t){\cal{V}}^{(2)}(\phi(t))
 \label{newfreq} \\
{\cal{V}}^{(1)}_{\vec{k}}(\phi(t))
     & = & {\cal{V}}^{(1)}(\phi(t))
\sqrt{\Omega}\delta_{\vec{k},0}
\end{eqnarray}
The ansatz for the Gaussian density matrix is the same as before
(\ref{densitymatrix}), as are the evolution equations for the coefficients
$A_k(t), \ B_k(t), \ {\cal{N}}_k(t)$.
However, the frequencies are now given by (\ref{newfreq}). The {\it classical}
equations of motion (\ref{pieq},\ref{fieq}) now become
\begin{eqnarray}
-\dot{\pi}                     & = & {\cal{V}}^{(1)}(\phi(t))a^3(t)
 \label{newpieq} \\
\dot{\phi}                     & = & \frac{\pi}{a^3(t)}
\label{newfieq}
\end{eqnarray}
The equations for the coefficients $A_k(t), \  B_k(t), \ {\cal{N}}_k(t)$ are
again solved in terms of the mode functions given in  appendix A but with
$V''(\phi_{cl}(t))$ now replaced by ${\cal{V}}^{(2)}(\phi(t))$ and the
following
replacement of the adiabatic frequencies:

\begin{eqnarray}
 W_k(t_o) \rightarrow {\cal{W}}_k(t_o) & = &
 {{\left[ {\vec{k}^2}+
{m^2(T_o)} \right]^{\frac{1}{2}}}\over {a(t_o)}}
\label{hartfreq} \\
\frac{m^2(T_o)}{a^2(t_o)}              & = & {\cal{V}}^{(2)}(\phi(t_o))
\label{defmass}
\end{eqnarray}

The mode functions of the appendix obey the differential equations and initial
conditions with this replacement and \[\mid \varphi_k(t) \mid^2 \rightarrow
\mid \varphi^H_k(t) \mid^2 .\] The
initial conditions at $t_o$ are now given in terms of these new adiabatic
frequencies (see appendix A).  The equal time two-point function thus becomes
\begin{equation}
\langle \eta^2(\vec{x},t) \rangle =  \frac{\hbar}{2}
 \int \frac{d^3k}{(2\pi)^3}
\mid \varphi^H_k(t) \mid^2
\coth\left[\beta_o\hbar {\cal{W}}_k(t_o)/2
\right], \label{newtwopointfunc}
\end{equation}
which leads to the following set of self-consistent time dependent
Hartree equations:
\begin{eqnarray}
& & \ddot{\phi}+3 \frac{\dot{a}}{a}\dot{\phi}+V'(\phi)+
\lambda \phi \frac{\hbar}{2}  \int \frac{d^3k}{(2\pi)^3}
\frac{\mid \varphi^H_k(t) \mid^2}{2}
\coth\left[\beta_o\hbar {\cal{W}}_k(t_o)/2
\right] = 0 \label{harteqofmot} \\
& & \left[\frac{d^2}{dt^2}+3 \frac{\dot{a}(t)}{a(t)}\frac{d}{dt}+
\frac{\vec{k}^2}{a^2(t)}+V''(\phi(t)) + \right. \\
& & \left. \lambda \frac{\hbar}{2}
 \int \frac{d^3k}{(2\pi)^3}
\frac{\mid \varphi^H_k(t) \mid^2}{2}
\coth\left[\beta_o\hbar {\cal{W}}_k(t_o)/2
\right] \right]
\varphi^H_k(t)
=0 \label{newdiffeqU} \\
& &
\varphi^H_k(t_o)        =  \frac{1}{\sqrt{a^3(t_o){\cal{W}}_k(t_o)}}
\label{newboundconU} \\
& &
\dot{\varphi}^H_k(t)\mid_{t_o}  =
 i\sqrt{\frac{{\cal{W}}_k(t_o)}{a^3(t_o)}}
\label{newboundconUdot}
\end{eqnarray}

 The renormalization aspects of these equations have been worked
out in detail in reference\cite{boyfrw} to which we refer the reader.

We find\cite{boyfrw} that the renormalized self-consistent Hartree equations
become
after letting $\Lambda = \infty$:
\begin{eqnarray}
& & \ddot{\phi}+3 \frac{\dot{a}}{a}\dot{\phi}+m^2_R\phi+\xi_R{\cal{R}}
\phi+ \frac{\lambda_R}{2} \phi^3+
\frac{\lambda_R \phi}{2} \langle \eta^2 \rangle_R =0
 \label{renharteqofmot} \\
& & \left[\frac{d^2}{dt^2}+3 \frac{\dot{a}(t)}{a(t)}\frac{d}{dt}+
\frac{\vec{k}^2}{a^2(t)}+m^2_R+ \xi_R {\cal{R}}+ \frac{\lambda_R}{2}
\phi^2 +
 \frac{\lambda_R}{2} \langle \eta^2 \rangle_R \right]
 \varphi^H_k(t) =0 \label{rennewdiffeqU}
\end{eqnarray}
where $\langle \eta^2 \rangle_R$ is the renormalized composite
operator\cite{boyfrw}.

For completeness we quote the renormalized equation of motion for the
order  parameter {\it to one-loop}

\begin{eqnarray}
& & \ddot{\phi}+3 \frac{\dot{a}}{a}\dot{\phi}+m^2_R\phi+\xi_R{\cal{R}}
\phi+ \frac{\lambda_R}{2} \phi^3+
\frac{\lambda_R \phi}{2} \langle \eta^2 \rangle_R  = 0
  \label{1lupreneqofmot} \\
& & \left[\frac{d^2}{dt^2}+3\frac{\dot{a}(t)}{a(t)}\frac{d}{dt}+
\frac{\vec{k}^2}{a^2(t)}+m^2_R+ \xi_R {\cal{R}}+ \frac{\lambda_R}{2}
\phi^2(t)  \right]
 \varphi_k(t) =0 \label{1luprennewdiffeqU} \\
\end{eqnarray}

The one-loop coupling constant renormalization differs from that in the Hartree
approximation by a factor of three. This is a consequence of the fact that the
Hartree approximation is equivalent to a large N approximation and only sums up
the s-channel bubbles.

\subsection{High Temperature Limit}

One of the payoffs of understanding the renormalization aspects and the
large momentum behaviour of the various fluctuation integrals is that
that it permits the evaluation of the high temperature limit. We shall
perform our analysis of the high temperature expansion for the Hartree
approximation. The one-loop case may be read off from these results.

The finite temperature contribution is determined by the integral

\begin{equation}
J =
\hbar
\int \frac{d^3k}{(2\pi)^3}
\frac{\mid \varphi^H_k(t) \mid^2}{e^{\beta_o
\hbar {\cal{W}}_k(t_o)}-1}  \label{tempint}
\end{equation}
For large temperature, only momenta $k \geq T_o$ contribute.
The
leading contribution is found to be\cite{boyfrw}
\begin{equation}
J = \frac{1}{12 \hbar}
\left[ \frac{k_B T_o a(t_o)}{a(t)}\right]^2
\left[1+{\cal{O}}(1/T_o) + \cdots \right] \label{highT}
\end{equation}

Thus we see that the leading high temperature behavior reflects the physical
red-shift in the cosmological background and it results in an effective time
dependent temperature

\[ T_{eff}(t) = T_o \left[\frac{a(t_o)}{a(t)}\right] \]

To leading order, the expression obtained for the time dependent effective
temperature corresponds to what would be obtained for an {\it adiabatic}
(isentropic) expansion for blackbody-type radiation consisting of massless
relativistic particles evolving in the cosmological background.

This behavior only appears at {\it leading} order in the high temperature
expansion. There are subleading terms that must be taken into account. These
can be calculated within the high temperature expansion and we do this below.
To avoid cluttering of notation, we will set $k_B=\hbar=1$ in what follows.
We find the linear and logarithmic contributions in $T_o$ to be\cite{boyfrw}

\begin{equation}
J_{1lin} = \left[\frac{a(t_o)}{a(t)}\right]^3 \frac{T_o m(T_o)}{\pi^2}
\int^{\infty}_{1} dz \left\{a^3(t) m(T_o)\sqrt{z^2-1}\frac{\mid \varphi^H_k(t)
\mid^2}{2}-
\frac{a(t)}{2a(t_o)} \right\}
\label{linearT}
\end{equation}
Note that the above integral is finite.

We obtain after some straightforward algebra\cite{boyfrw}:

\begin{equation}
J_{1log} = -\frac{\ln\left[\frac{m(T_o)}{T_o}\right]}{8\pi^2}
\left\{-\frac{{\cal{R}}}{6}-\frac{\dot{a}^2(t_o)}{a^2(t)}+
\left[\frac{a(t_o)}{a(t)}\right]^2\left[m^2(T_o)+
\frac{\lambda_R T^2_c}{24}\right] -\frac{\lambda_R T^2_c}{24}\right\}
\end{equation}
That is, in the limit $T_o >> m(T_o),~ J_1 = J_{1lin} + J_{1log} +
O((T_o)^0)$.

Comparing the ${\cal{O}}(T_o^2, T_o, \ln T_o)$ contributions it becomes
clear that they have very different time dependences through the
scale factor $a(t)$. Thus the high temperature expansion as presented
will not remain accurate at large times since the term quadratic in $T_o$
may become of the same order or smaller than the linear or logarithmic terms.
The high temperature expansion and the long time limit are thus not
interchangeable, and any high temperature expansion is thus bound to
be valid only within some time regime that depends on the initial
value of the temperature and the initial conditions.

As an illustration of this observation, we calculate $J_{1lin}$ explicitly in
the case of de Sitter space.  We need to obtain $\mid \varphi^H_k(t) \mid^2$
in order to evaluate the integral in (\ref{linearT}). Inserting the term
proportional to
$T_o^2$ in the Hartree equations, we find that $\mid \varphi^H_k(t) \mid^2$
obeys the differential equation
\begin{equation}
\left[ \frac{d^2}{dt^2}+3H\frac{d}{dt}+\left[\frac{k^2}{a_o^2}+m^2(T_o)+
\frac{\lambda_R T^2_c}{24}\right]e^{-2Ht}-\frac{\lambda_R T^2_c}{24}
\right]\varphi^H_k(t) = 0 \label{desit}
\end{equation}
The solution of this equation is given by:
\begin{eqnarray}
\varphi^H_k(t) & = & \left[ C_1 H^{(1)}_{3/2}(B_ke^{-Ht})+
C_2 H^{(2)}_{3/2}(B_ke^{-Ht}) \right] e^{-\frac{3}{2}Ht}
\label{sols} \\
     B_k & = & \frac{1}{H}\left[\frac{k^2}{a_o^2}+m^2(T_o)+
\frac{\lambda_R T^2_c}{24}\right]^{\frac{1}{2}} \nonumber
\end{eqnarray}
where $H_{3/2}^{(1,2)}$ are the Hankel functions and we have assumed
\[m^2(T_o)\; ; \; \frac{\lambda_R T^2_c}{24} \ll H . \]

The coefficients $C_1 , C_2$ are determined by the initial conditions
on $\varphi^H_k(t)$ described above.  We finally obtain
\begin{equation}
J_{1lin} = \frac{m(T_o)T_o}{8\pi^2}\left(H e^{Ht_o}\right)^4
\int^{\infty}_{1}
\frac{dz}{z} \frac{\sqrt{z^2-1}}{\left[\frac{\lambda_R T^2_c}{24}+m^2(T)z
\right]^2}
\end{equation}
This term is time independent, finite and positive. This example clearly
illustrates the fact that different powers of $T_o$ enter in the expansion
with different functions of time and that the high temperature expansion
is non-uniform as a function of time.

\subsection{Evolution of the Initial Distribution}

The initial density matrix at $t=t_o$ was assumed to be thermal for the
adiabatic modes. This corresponds to a Boltzmann distribution for the
uncoupled harmonic oscillators for the adiabatic modes of momentum $\vec{k}$,
and frequencies ${\cal{W}}_k(t_o)$ (in the Hartree approximation; as usual,
the one-loop result can be found by replacing this with $W_k(t_o)$). That is
\begin{eqnarray}
\rho(t_o) & = &  \frac{e^{-\beta_o H_o}}{Tr e^{-\beta_o H_o}}
\label{diagdensmat} \\
H_o       & = & \sum_{\vec{k}}\hbar {\cal{W}}_k(t_o)\left[
\alpha^{\dagger}_k(t_o)\alpha_k(t_o)+\frac{1}{2}\right] \label{iniham}
\end{eqnarray}
\noindent The creation and annihilation operators  define the initial
occupation
number of the adiabatic modes:
\begin{eqnarray}
\hat{N}_k(t_o)                 & = &
\alpha^{\dagger}_k(t_o)\alpha_k(t_o) \label{numberop}\\
\langle \hat{N}_k(t_o) \rangle & = &
\frac{1}{e^{\beta_o \hbar {\cal{W}}_k(t_o)}-1} \label{expecnum},
\end{eqnarray}
where the expectation value in (\ref{expecnum}) is in the initial density
matrix at time $t_o$.

In a time dependent gravitational background the concept of particle is
ill-defined. However, by postulating an equilibrium initial density matrix of
the above form, a preferred ``pointer'' basis is singled out at the initial
time. It is this basis that provides a natural definition of particles  at the
initial time and we can use it to ask: how does the expectation value of {\it
this} number operator evolve in time?

At any time $t$, this expectation value is given by
\begin{equation}
\langle \hat{N}_k \rangle (t) = \frac{Tr\left[
\alpha^{\dagger}_k(t_o)\alpha_k(t_o) \rho(t)\right]}{Tr \rho(t_o)}
\label{timedepnum}
\end{equation}
This quantity gives information on how the {\em original} Boltzmann
distribution function for the adiabatic modes evolves with time.
The $\vec{k} = 0$
mode will receive a contribution from the order parameter, but since the number
of particles is not conserved (no charge) there is no bose condensation and the
$\vec{k} =0$ mode will give a negligible contribution to the total number of
particles. Thus we only concentrate on the $\vec{k} \neq 0$ modes.

The expectation value (\ref{timedepnum}) may be easily computed by writing the
creation and annihilation operators in terms of $\eta_k,\ \Pi_k= \delta \slash
\delta \eta_{-k}$ in the {Schr\"{o}dinger} picture at $t_o$.
The result of doing this is:

\begin{eqnarray}
\alpha^{\dagger}_k(t_o) & = & \frac{1}{\sqrt{2\hbar}}\left[
-\frac{1}{\sqrt{a^3(t_o){\cal{W}}_k(t_o)}}\frac{\delta}{\delta \eta_k}
+\sqrt{a^3(t_o){\cal{W}}_k(t_o)}\eta_{-k} \right] \label{crea} \\
\alpha_k(t_o)           & = & \frac{1}{\sqrt{2\hbar}}\left[
\frac{1}{\sqrt{a^3(t_o){\cal{W}}_k(t_o)}}\frac{\delta}{\delta \eta_{-k}}
+\sqrt{a^3(t_o){\cal{W}}_k(t_o)}\eta_{k} \right] \label{dest}
\end{eqnarray}

After some straightforward algebra we find
\begin{eqnarray}
\langle \hat{N}_k \rangle (t)+ \frac{1}{2} & = &
\left(2 \mid {\cal{F}}_k(t,t_o) \mid^2 -1 \right)
\left( \langle \hat{N}_k \rangle (t_o)+ \frac{1}{2} \right)
\label{numrel} \\
\mid {\cal{F}}_k(t,t_o) \mid^2             & = &
\frac{1}{4}\frac{\mid \varphi^H_k(t) \mid^2}{\mid \varphi^H_k(0) \mid^2}
 \left[1+ \frac{a^6(t)}{a^6(t_o){\cal{W}}^2_k(t_o)}
\frac{\mid{\dot{\varphi}}^H_k(t)\mid^2}{\mid \varphi^H_k(t)
\mid^2}\right] + \frac{1}{2}
 \label{bogol}
\end{eqnarray}

where we have made use of (\ref{arealfin}, \ref{aimagfin}). This result
exhibits the two contributions from ``spontaneous'' (proportional to the
initial thermal occupation) and ``induced'' (independent of it).

We now show that this result may be understood as a Bogoliubov
transformation. To do this, consider the expansion of the field in the
{\it Heisenberg} picture:
\begin{equation}
\eta_k(t) = \frac{1}{\sqrt{2}} \left[ \tilde{\alpha}_k
 \varphi^{H\dagger}_k(t)+
{\tilde{\alpha}}^{\dagger}_k \varphi^H_k(t) \right] \label{heisfield}
\end{equation}
where the mode functions satisfy
\begin{equation}
\frac{d^2 \varphi^H_k}{dt^2} + 3\frac{\dot{a}}{a}\frac{d\varphi^H_k}{dt}+
\left[\frac{\vec{k}^2}{a^2}+{\cal{V}}^{(2)}(\phi(t))\right]\varphi^H_k =0
\label{heishart}
\end{equation}
together with the self-consistency relation.
Then the Heisenberg field $\eta_k(t)$ is a solution of the Hartree
Heisenberg equations of motion and the $\tilde{\alpha}^{\dagger}_k \; ,
\tilde{\alpha}_k$ create and destroy the Hartree-Fock states. Notice that in
the Heisenberg picture, these creation and annihilation operators {\it
do not depend on time}. Using the Wronskian properties of the functions
$\varphi^H_k(t)$ (see appendix A) we can invert and find the creation and
annihilation operators in terms of $\eta_k(t)$ and its canonically
 conjugate momentum $\Pi_{-k}(t)$. Once we have expressed
 these operators in the
 Heisenberg picture in terms of the field and its canonically conjugate
momentum, we can go to the {Schr\"{o}dinger} picture at time $t_o$.
In this picture the creation and annihilation operators
 depend on time and are given by
\begin{eqnarray}
\tilde{\alpha}_k (t) & = & \frac{i}{\sqrt{2}}\left[\Pi_k(t_o)
\varphi^H_k(t)-a^3(t)\eta_k(t_o)\dot{\varphi}^H_k(t) \right]
 \label{schpicdag} \\
{\tilde{\alpha}}^{\dagger}_{-k} (t)
      & = &  \frac{-i}{\sqrt{2}}\left[\Pi_k(t_o)
\varphi^{H\dagger}_k(t)-a^3(t)\eta_k(t_o)\dot{\varphi}^{H\dagger}_k(t)\right]
 \label{schpic}
\end{eqnarray}

The {Schr\"{o}dinger} picture fields at $t_o$ can be written in terms of the
operators (\ref{crea}, \ref{dest}) and we finally find the creation and
destruction operators at time $t$ to be related to those at time $t_o$
by a Bogoliubov transformation:
\begin{equation}
\tilde{\alpha}_k (t) = {\cal{F}}_{+,k}(t,t_o) {\alpha}_k(t_o) +
{\cal{F}}_{-,k}(t,t_o)
{\alpha}^{\dagger}_{-k} (t_o) \label{bogoltrans}
\end{equation}
If we now compute the average of the new creation and annihilation operators in
the initial density matrix and write the mode functions $\varphi_k(t)$ in terms
of the real functions ${\cal{U}}_{1,2}$ as defined in  appendix A, we recognize
$\mid {\cal{F}}_{+,k}(t,t_o) \mid^2 $ to be the same as $\mid
{\cal{F}}_{k}(t,t_o) \mid^2$ given by (\ref{bogol}).

We also find that
\[\mid {\cal{F}}_{+,k}(t,t_o) \mid^2 -
 \mid {\cal{F}}_{-,k}(t,t_o) \mid^2 = 1 \]
as is required for a Bogoliubov transformation.

One way to interpret this result is that, at least within the one-loop or
Hartree approximations, time evolution corresponds to a Bogoliubov
transformation. This interpretation is, in fact, consistent with
the result that in these approximation schemes, the density matrix
remains Gaussian with the only change being that the covariance and
mixing terms change with time.

Thus within the one-loop or Hartree approximation, time evolution corresponds
to a ``squeezing'' of the initial state. The covariance changes with time and
this corresponds to a Bogoliubov transformation.
As argued by Grishchuk and Sidorov\cite{grishchuk} the amplification of quantum
fluctuations  during inflation is a process of quantum squeezing and it
corresponds to a Bogoliubov transformation.  The properties of these
``squeezed'' quantum states have been investigated in references
\cite{branden,gasperini,albrecht}.

Hu and Pavon\cite{hupavon}, Hu and Kandrup\cite{hukandrup} and
Kandrup\cite{kandrup} have introduced a non-equilibrium, coarse-grained entropy
that grows in time as a consequence of particle production and ``parametric
amplification''. This definition was generalized by Brandenberger et
al. \cite{branden}, and Gasperini and  collaborators \cite{gasperini} to give a
measure of the entropy of the gravitational field. The growth of this entropy
is again a consequence of the parametric amplification of fluctuations and the
``squeezing'' of the quantum state under time evolution.

These authors argue that the non-equilibrium coarse-grained entropy in the
mode of (comoving) wavevector $\vec{k}$ is
\begin{equation}
s_k \approx \ln\left(\langle N_k \rangle (t)\right)
\label{entropy}
\end{equation}
in the case when $\langle N_k \rangle (t) \gg 1$.
Thus the growth of entropy is directly associated with ``particle
production'' or in our case to the evolution of the initial Boltzmann
distribution function.

The coefficient of parametric amplification is related to the Bogoliubov
coefficient  given by equation
(\ref{bogol}). Thus this coefficient directly determines the time dependence
of the non-equilibrium coarse-grained entropy.

\subsection{Appendix A}

The Riccati equation (\ref{riccati}) can be transformed into a linear
differential equation by the change of variables
\begin{equation}
{\cal{A}}_k(t) = -ia^3(t)\frac{\dot{\varphi}_k(t)}{\varphi_k(t)}
\label{fiofk}
\end{equation}
We find that $\varphi_k(t)$ obeys a simple evolution equation
\begin{equation}
\frac{d^2 \varphi_k}{dt^2} + 3\frac{\dot{a}}{a}\frac{d\varphi_k}{dt}+
\left[\frac{\vec{k}^2}{a^2}+V''(\phi_{cl}(t))\right]\varphi_k =0
\label{fievolution}
\end{equation}
with $\phi_{cl}(t)$ the solution to the {\it classical} equation of
motion (\ref{pieq}, \ref{fieq}).
In the Hartree approximation $V''(\phi_{cl}(t))$ should be replaced
by ${\cal{V}}^{(2)}(\phi(t))$ with $\phi(t)$ the {\it full solution}
of the self-consistent equations.
\begin{equation}
\frac{d^2 \varphi^H_k}{dt^2} + 3\frac{\dot{a}}{a}\frac{d\varphi^H_k}{dt}+
\left[\frac{\vec{k}^2}{a^2}+{\cal{V}}^{(2)}(\phi(t))\right]\varphi^H_k =0
\label{fihevolution}
\end{equation}

The Wronskian for two arbitrary solutions to the above differential
equation is
\begin{equation}
{\bf{W}}[\varphi_1,\varphi_2] = \dot{\varphi_2}(t)\varphi_1(t)
-\dot{\varphi_1}(t)\varphi_2(t) = \frac{C}{a^3(t)} \label{wronskian}
\end{equation}
with $C$ a constant. Then writing $\varphi_k(t) = \varphi_{1k}(t)+i
\varphi_{2k}(t)$ with $\varphi_{1,2}$ {\it real solutions} we find
\begin{eqnarray}
{\cal{A}}_{Rk}(t) & = & \frac{C}{\varphi^2_{1k}+\varphi^2_{2k}}
\label{area} \\
{\cal{A}}_{Ik}(t) & = & -a^3(t)\;\frac{\dot{\varphi_{1k}}\varphi_{1k}+
\dot{\varphi_{2k}}\varphi_{2k}}{\varphi^2_{1k}+\varphi^2_{2k}}
\label{aima}
\end{eqnarray}

 It proves convenient to introduce (for all $\vec{k}$)
 the real functions ${\cal{U}}_{1,2}(t)$
as
\begin{equation}
\varphi_{1,2}(t) = [a(t)]^{-\frac{3}{2}}
\frac{{\cal{U}}_{1,2}(t)}{\sqrt{W_k(t_o)}}
 \label{calU}
\end{equation}
The ${\cal{U}}_{\alpha k}$ for $\alpha=1,2$ are real and
satisfy the {Schr\"{o}dinger}-like differential equation
\begin{eqnarray}
& & \left[\frac{d^2}{dt^2}-\frac{3}{2}\left(\frac{\ddot{a}}{a}+\frac{1}{2}
\frac{\dot{a}^2}{a^2}\right)+\frac{\vec{k}^2}{a^2(t)}+
V''(\phi_{cl})(t)\right]
{\cal{U}}_{\alpha k}(t)
=0 \label{diffeqU} \\
& & {\bf{W}}[{\cal{U}}_{1,k},{\cal{U}}_{2,k}]= C W_k(t_o)
\label{wronskchi}
\end{eqnarray}
with ${\bf{W}}[\cdots]$ the Wronskian, and $C$ is
the same constant as above.
Since the choice of $C$ corresponds to a choice of normalization of
these functions, we choose $C=1$. The initial conditions (\ref{Are},
\ref{Aim}) still leave one free condition on these functions, we
choose
\begin{eqnarray}
{\cal{U}}_{1k}(t_o) & \neq 0 & \label{calU1} \\
{\cal{U}}_{2k}(t_o) &  =     & 0 \label{calU2}
\end{eqnarray}

The boundary conditions on the mode functions ${\cal{U}}_{\alpha,k}$
are
\begin{eqnarray}
& & {\cal{U}}_{1k}(t_o) = 1 \; \; ; \; \; {\cal{U}}_{2k}(t_o) = 0
\label{boundconU} \\
& & \dot{{\cal{U}}}_{1k}(t_o) = \frac{3}{2}\frac{\dot{a}(t_o)}{a(t_o)}
\; \; ; \; \; \dot{{\cal{U}}}_{2k}(t_o) = W_k(t_o) =
\left[\frac{\vec{k}^2}{a^2(t_o)}+V''(\phi_{cl}(t_o))\right]^{\frac{1}{2}}
\label{boundconUdot}
\end{eqnarray}
and the corresponding replacement for the Hartree case.
Thus the final solution to the Riccati equation (\ref{riccati}) with
the given initial conditions is
\begin{eqnarray}
{\cal{A}}_{Rk}(t) & = & \frac{a^3(t)W_k(t_o)}{\left[{\cal{U}}^2_{1k}(t)+
{\cal{U}}^2_{2k}(t)\right]}
\label{arealfin} \\
{\cal{A}}_{Ik}(t) & = & -a^3(t)\left[ \frac{
{\cal{U}}_{1k}\left(\dot{{\cal{U}}}_{1k}
-\frac{3\dot{a}}{2a}{\cal{U}}_{1k}\right)+
{\cal{U}}_{2k}\left(\dot{{\cal{U}}}_{2k}
-\frac{3\dot{a}}{2a}{\cal{U}}_{2k}\right)}
{ {\cal{U}}^2_{1k}(t)+ {\cal{U}}^2_{2k}(t) }  \right] \label{aimagfin}
\end{eqnarray}

In terms of the original functions $\varphi_k(t)$ (\ref{fiofk}) the
initial conditions are simply
\begin{eqnarray}
\varphi_k(t_o)               & = & \frac{1}{\sqrt{a^3(t_o)W_k(t_o)}}
 \label{bounfi} \\
\dot{\varphi}_k(t)\mid_{t_o} & = &
 i \sqrt{\frac{W_k(t_o)}{a^3(t_o)}}
\label{bounfidot}
\end{eqnarray}

Then the initial conditions for $\varphi_k(t) \; ; \; {\varphi}^*_k(t)$
are naturally interpreted as those for negative and positive (adiabatic)
frequency modes at the initial time $t_o$.

The  reason for introducing the functions ${\cal{U}}_{\alpha,k}(t)$
is because these obey a simpler second order Schr\"{o}dinger-like equation
which is amenable to be studied in the asymptotic regime via WKB
approximations (see the section on renormalization).

\subsection{Appendix B: Conformal Time Analysis}

It is interesting to see how some of our results can be obtained by rewriting
the metric in terms of the conformal time defined by:
\begin{equation}
\eta = {\int}^{t} \frac{dt'}{a(t')}
\end{equation}

The first thing we should note is that the physics should not depend on what
time coordinate is used, since the theory should be generally coordinate
invariant (there are no gravitational anomalies in four
 dimensions). Thus, the field amplitude and the canonical momentum in
conformal time should be related to those in comoving time via a canonical
transformation. We now show that this is indeed the case.

Using the conformal time version of the line element $ds^2 = a^2(\eta)(d\eta^2
- d\vec{x}^2)$, the scalar field action can be written as:

\begin{equation}
S[\Phi] = \int d\eta d^3x a^4(\eta) \left[ \frac{1}{2 a^2(\eta)}
((\partial_{\eta}\Phi)^2 - (\nabla \Phi)^2) -
V(\Phi(\eta, \vec{x}))\right],
\end{equation}
where the potential term is as in the text (i.e. it could include a coupling to
the curvature scalar). Following the standard procedure for obtaining the
canonical momentum $\Pi(\eta, \vec{x})$ to $\Phi(\eta, \vec{x})$, and
to get at the Hamiltonian density yields:

\begin{eqnarray}
\Pi(\eta, \vec{x}) & = & a^2(\eta) \Phi'(\eta, \vec{x})\nonumber \\
{\cal H} & = & \frac{\Pi^2}{2 a^2(\eta)} +
\frac{a^2(\eta)}{2} (\nabla \Phi)
+ a^4(\eta) V(\Phi).
\label{confham}
\end{eqnarray}
Here conformal time derivatives are denoted by a prime.
The generator $H$ of displacements in conformal time is the spatial integral of
${\cal H}$ above. Thus the conformal time Liouville equation reads:

\begin{equation}
i \frac{\partial \rho}{\partial \eta} = [H, \rho].
\end{equation}

If we label the field and its conjugate momemtum in the comoving time frame
(i.e. that of the text) as $\hat{\Phi},\ \hat{\Pi}$ respectively, the
results of section 2 are:

\begin{eqnarray}
\hat{\Pi}(\vec{x},t) & = & a^3(t)\dot{\hat{\Phi}}(\vec{x},t)\nonumber \\
\hat{H} & = &\int d^3x \left\{ \frac{\hat{\Pi}^2}{2a^3}+
\frac{a}{2}(\vec{\nabla}\hat{\Phi})^2+
a^3 V(\hat{\Phi}) \right\}
\end{eqnarray}

The Liouville equation in comoving time can be rewritten in conformal time
using the relation: $\partial \slash {\partial t} = a(\eta)^{-1}\partial \slash
{\partial \eta}$. After doing this we find that eq.(\ref{liouville}) becomes:
\begin{equation}
i\frac{\partial \hat{\rho}(\hat{\Pi}, \hat{\Phi})}{\partial \eta} =
\left[a(\eta) \hat{H}, \hat{\rho}(\hat{\Pi}, \hat{\Phi})\right].
\end{equation}
But
\begin{equation}
a(\eta) \hat{H} = \int d^3x \left[ \frac{\hat{\Pi}}{2 a^2(\eta)} + \frac{1}{2}
a^2(\eta) (\nabla \hat{\Phi})^2 + a^4(\eta) V(\hat{\Phi})\right].
\end{equation}
Comparing this with eq.(\ref{confham}), we see that we can make the
identifications: $\Pi=\hat{\Pi},\ \Psi= \hat{\Psi}$. Thus not only is the
physics equivalent in both coordinate systems (as must have been the case), but
the physics in the two coordinate systems related by a trivial canonical
transformation.

We can rewrite all of the comoving time results in terms of conformal time.
Some of the equations take on a much simpler form in conformal time than in
comoving time. This will be important when numerical issues are tackled, such
as an analysis of the back-reaction problem. This work is in progress.

\section{Dynamics of Chiral Simmetry Breaking: Disoriented Chiral Condensates}

The proposition has been put forth recently that regions of misaligned chiral
vacuum, or  disoriented chiral condensates (DCC) might form in either
ultra-high energy or heavy nuclei collisions\cite{dcc1}-\cite{dcc4}.  If so,
then this would be  a striking probe of the QCD phase transition. It might also
help  explain\cite{dcc5,dcc6}  the so-called Centauro and anti-Centauro events
observed in high-energy cosmic ray experiments\cite{dcc7}.

How can we tell, from a theoretical standpoint, whether or not we should expect
a DCC to form? Clearly, investigating QCD directly is out of the question  for
now; the technology required to compute the evolution of the relevant  order
parameters directly from QCD is still lacking, though we can use lattice
calculations for hints about some aspects of the QCD phase transition.  What we
need then is a model that encodes the relevant aspects of QCD in a  faithful
manner, yet is easier to calculate with than QCD itself.

Wilczek and Rajagopal \cite{wilraj1} have argued  that the $O(4)$ linear
$\sigma$-model is such a model.  It lies within the same  {\em static}
universality class as QCD with {\em two} massless quarks,  which is a fair
approximation to the world at temperatures and energies below $\Lambda_{QCD}$.
Thus work done on the $n=4$ Heisenberg ferromagnet can  be used to understand
various static quantities arising at the chiral phase transition.

One conclusion from reference \cite{wilraj1} was that, if as the critical
temperature for the chiral phase transition was approached from above the
system remained in thermal equilibrium, then it was very unlikely that a large
DCC region, with its concomitant biased pion emission, would form.

 The point was that the correlation length $\xi=m_{\pi}^{-1}$
 did {\em not} get large  compared to the $T_c^{-1}$. A more quantitative
criterion involving the comparison of
the energy in a correlation volume just below $T_c$ with the $T = 0$ pion mass
(so as to find the number of pions in a correlation volume) supports the
conclusion that as long as the system can equilibrate, no large regions of
DCC will form.

The only option left, if we want to form a DCC, is to insure that the system
is far out of equilibrium. This can be achieved by {\em quenching} the
 system (although Gavin and M\"{u}ller\cite{gavin} claim that {\em annealing}
 might also work). What this means is the following.
 Start with the system in equilibrium at a
temperature above $T_c$. Then suddenly drop the temperature to zero. If the
rate at which the temperature drops is much faster than the rate at which the
system can adapt to this change, then the state of the system after the quench
is such that it is still in the thermal state at the initial temperature.
However, the {\em dynamics} governing the evolution of that initial state is
now driven by the $T=0$ Hamiltonian. The system will then have to relax from
the initial state, which is {\em not} the ground state of the Hamiltonian to
the zero temperature ground state. During this time, it is expected that
regions in which the order parameter is correlated will grow. We can then hope
that the correlation regions will grow to be large enough to contain a large
number of pions inside them.

The possibility that the chiral phase transition might occur following
a  quench in heavy-ion collisions was explored by Wilczek and Rajagopal in
reference\cite{wilraj2}.They argue there that long wavelength fluctuations in
the pion fields can develop after the quench occurs. Modes with wavenumbers
 $k$ smaller than some critical wavenumber $k_{\rm{crit}}$ will be unstable
 and regions in which the pion field is correlated will grow in spatial
 extent for a period of time. The essence of this mechanism is that
the pions are the
would-be Goldstone particles of spontaneous chiral symmetry breaking. In
the absence of quark masses, the pions would be exactly massless when the
$\vec{\pi}$ and $\sigma$ fields are in their ground state. However, during the
quench, the $\sigma$ field is displaced from its zero temperature minimum so
that the required cancellation between the negative bare $\rm{mass}^2$ term
in the Lagrangian and the $\rm{mass}^2$ induced through the pion interactions
with the $\sigma$ condensate does not occur. This then allows some of the pion
momentum modes to propagate as if they had a {\em negative} $\rm{mass}^2$,
thus causing exponential growth in these modes.

Some studies were already reported in which the classical evolution
equations of the linear sigma model were studied analytically and
numerically \cite{wilraj2,pisarski}.

In this section we use the techniques developed in the previous section to
study the non-equilibrium dynamics of the phase transition.
In particular we will focus on the formation and growth of domains of
correlated pions to assess the possibility of large domains leading to
the production of a large number of correlated pions that may be detected
during energetic $P-\bar{P}$ or heavy ion collisions.

\subsection{\bf The $O(4)$ $\sigma$-Model Out of Equilibrium}

Our strategy is as follows. We will use the techniques developed in
the
previous
section based on  the functional
Schr\"{o}edinger representation,
in which the time evolution of the system is represented by the
 time evolution of its {\em density matrix}.

The next step is to evolve the density matrix in time from this initial
 state via the quantum Liouville equation:

\be
i\h \frac{\partial \rho (t)}{\partial t} = [H, \rho(t)],
\ee
where $H$ is the Hamiltonian of the system {\em after} the quench. Using this
density matrix, we can, at least in principle, evaluate the equal time
correlation function for the pion fields, and observe its growth with time.

Let us now implement this procedure. We start with the sigma model Lagrangian
density

\ba
{\cal L} & = & \frac{1}{2}\partial_{\mu}\vec{\Phi}\cdot
\partial^{\mu}\vec{\Phi}
- V(\sigma, \vp) \\
V(\sigma, \vp)
         & = & \frac{1}{2} m^2(t) \vP \cdot \vP + \lambda
(\vP \cdot \vP)^2 - h \sigma \label{potente}
\ea
where $\vP$ is an $O(N+1)$ vector,
$\vP= (\sigma, \vp)$ and $\vp$ represents
the $N$ pions.

The linear sigma model is a low energy effective theory for an
$SU_{\rm{L}}(2)\times SU_{\rm{R}}(2)$ (up and down quarks) strongly interacting
theory. It may be obtained as a Landau-Ginzburg effective theory from a
Nambu-Jona-Lasinio model\cite{klebansky}.
In fact, Bedaque and Das\cite{bedaque} have studied a quench starting
from an $SU_{\rm{L}}(2)\times SU_{\rm{R}}(2)$ Nambu-Jona-Lasinio model.

We have parametrized the dynamics of the cooling down process in terms of a
time dependent mass term. We can use this to describe the phenomenology
 of either a sudden quench where the mass$^2$ changes sign
instanteneously or that of a relaxational
process in which the mass$^2$ changes sign on a time scale determined by the
dynamics. In a heavy ion collision, we expect this relaxation time
scale to be of the order of $\tau \sim 0.5-1 \rm{fm/c}$.
The term $h \sigma$ accounts for the explicit breaking of chiral symmetry  due
to the (small) quark masses. We leave $N$ arbitrary for now, though at
 the end we will take $N=3$.

Furthermore we will study the dynamics in real Minkowsky time rather than
the original rapidity variables as first proposed by Bjorken. Recently
Cooper et. al.\cite{coopmoto} have reported on an investigation
of the dynamics using the rapidity variables with very similar results.

 Our first order of business is to identify the correct order
parameter for the phase transition and then to obtain its equation of motion.
Let us define the fluctuation field operator $ \chi(\x,t)$ as

\be
\sigma = \phi(t) + \chi(\x,t),
\ee
with $\phi(t)$ a c-number field defined by:

\ba
\phi(t) & = & \frac{1}{\Omega} \int d^3 x \langle \sigma(\x) \rangle
 \nonumber \\
        & = & \frac{1}{\Omega} \int d^3 x \frac{{\rm{Tr}}(\rho(t)
\sigma(\x))}{{\rm{Tr}}\rho(t)} .
\ea
Here $\Omega$ is the spatial volume we enclose the system in. The fluctuation
field $\chi(\x,t)$ is defined so that (i) $\langle \chi(\x,t) \rangle=0$,
and
(ii) $\dot{\chi}(\x,t)=-\dot{\phi}(t)$. Making use of the Liouville equation
for
the density matrix, we arrive at the following equations:

\ba
\dot{\phi}(t) & = & p(t) = \frac{1}{\Omega} \int\ d^3 x\
\langle \Pi_{\sigma}(\x) \rangle \label{momexp} \\
\dot{p}(t)    & = & -\frac{1}{\Omega}\int d^3x\
\langle\frac{\delta V(\sigma, \vp)}{\delta \sigma(\vec{x})} \rangle ,
\ea
where $\Pi_{\sigma}(\x)$ is the canonical momentum conjugate to
 $\sigma(\x)$.

The derivative of the potential in the equation for $\dot{\pi}(t)$ is to
be evaluated at $\sigma = \phi(t) + \chi(\x,t)$. These equations can be
combined into a single one describing the evolution of the order parameter
$\phi(t)$:

\be
\ddot{\phi}(t) + \frac{1}{\Omega}\int d^3x\
\langle
\frac{\delta V(\sigma, \vp)}{\delta \sigma(\vec{x})}
\rangle
\mid_{\sigma = \phi(t) +
\chi(\x,t)}  = 0.
\ee

To proceed further we have to determine the density matrix. Since the
Liouville equation is first order in time we need only specify $\rho(t=0)$.
At this stage we could proceed to a perturbative description of the
dynamics (in a loop expansion).

However, as we learned previously in a
similar situation\cite{boyveg,boydom}, the non-equilibrium
dynamics of the phase transition cannot be studied within perturbation theory.

Furthermore,
since the quartic coupling of the linear sigma model $\lambda$
must be large ($\lambda \approx 4-5$ so as to reproduce the value of $f_{\pi}
\approx 95$ Mev with a ``sigma mass'' $\approx$ 600 Mev), the linear
sigma model is a {\it strongly} coupled theory and any type
 of perturbative expansion will clearly be unreliable.
Thus,  following our previous work\cite{boydom,boyfrw} and
the work of Rajagopal and Wilczek\cite{wilraj2} and Pisarski\cite{pisarski}
 we invoke a Hartree approximation.

In the presence of a vacuum expectation value, the Hartree factorization is
somewhat subtle. We will make a series of {\it assumptions} that we feel are
quite reasonable but which, of course, may fail to hold under some
circumstances and for which we do not have an {\it a priori} justification.
These are the following: i) no cross correlations between the pions and the
sigma field, and ii) that the two point correlation functions of the pions are
diagonal in isospin space, where by isospin we now refer to the unbroken
$O(N)\ (N=3)$
symmetry under which the pions transform as a triplet. These assumptions
lead to the following Hartree factorization of the non-linear terms in the
Hamiltonian:

\ba
\chi^4 & \rightarrow & 6 \langle \chi^2 \rangle \chi^2  +\mbox{constant}
\label{hartreechi4} \\
\chi^3 & \rightarrow & 3 \langle \chi^2 \rangle \chi
\label{hartreechi3} \\
(\vec{\pi}\cdot\vec{\pi})^2
       & \rightarrow & (2+\frac{4}{N})\langle \vec{\pi}^2 \rangle
\vec{\pi}^2 + \mbox{constant} \label{hartreepi4} \\
\vec{\pi}^2 \chi^2
       & \rightarrow & \vec{\pi}^2 \langle \chi^2 \rangle+ \langle
\vec{\pi}^2 \rangle \chi^2 \label{hartreepi2chi2} \\
\vec{\pi}^2 \chi
       & \rightarrow & \langle \vec{\pi}^2 \rangle \chi ,
\label{hartreepi2chi}
\ea
where by ``constant'' we mean the operator independent expectation
values of the composite operators. Although these will be present as
operator independent terms in the Hamiltonian, they are c-number terms
and will not enter in the time evolution of the density matrix.

It can be checked that when $\phi =0$ one obtains the
$O(N+1)$ invariant Hartree factorization.

In this approximation the resulting Hamiltonian is quadratic, with a
 linear term  in $\chi$:

\be
H_H(t) = \int d^3x \left\{ \frac{\Pi^2_{\chi}}{2}+
\frac{\vec{\Pi}^2_{\pi}}{2}+\frac{(\nabla \chi)^2}{2}+
\frac{(\nabla \vec{\pi})^2}{2}+\chi {\cal{V}}^1(t)+
\frac{{\cal{M}}^2_{\chi}(t)}{2}
\chi^2+\frac{{\cal{M}}^2_{\pi}(t)}{2}\vec{\pi}^2 \right\} .
 \label{hartreeham}
\ee

Here $\Pi_{\chi}, \ \vec{\Pi}_{\pi}$ are the canonical momenta
conjugate to $\chi(\x), \ \vec{\pi}(\x)$ respectively and
${\cal{V}}^1$ is recognized as the derivative of the Hartree
``effective potential''\cite{moshe,camelia} with respect to $\phi$
(it is the derivative of the non-gradient terms of the
effective action\cite{boydom,boyfrw}).

In the absence of an explicit symmetry breaking term, the Goldstone theorem
requires the existence of massless pions, ${\cal{M}}_{\pi}=0$ whenever
${\cal{V}}^1$ for $\phi \neq 0$.
However, this is
{\em not} the case within our approximation scheme as it stands.

This situation can be easily remedied, however,
by noting that the Hartree approximation becomes exact in the large
$N$
limit. In this limit,
 $\langle \vec{\pi}^2 \rangle
\approx {\cal{O}}(N), \ \langle \chi^2 \rangle \approx
 {\cal{O}}(1), \ \phi^2
\approx {\cal{O}}(N)$. Thus we will approximate further by neglecting the
${\cal{O}}(1/N)$ terms in the formal large
$N$
limit. This further truncation
ensures that the Ward identities are satisfied. We now obtain
\ba
{\cal{V}}^1(t)       & = & \phi(t) \left[m^2(t)+4\lambda \phi^2(t)+
4\lambda \langle \vec{\pi}^2 \rangle(t) \right] -h \label{nu1} \\
{\cal{M}}^2_{\pi}(t) & = & m^2(t)+4\lambda \phi^2(t)+4\lambda
\langle \vec{\pi}^2 \rangle (t) \label{pionmass} \\
{\cal{M}}^2_{\chi}(t)& = & m^2(t)+12\lambda \phi^2(t)+4\lambda
\langle \vec{\pi}^2 \rangle (t) . \label{chimass}
\ea
The Hamiltonian is now quadratic with time dependent self-consistent
masses and Goldstone's Ward identities are satisfied.

Since the evolution Hamiltonian is quadratic in this approximation, we
propose in the Hartree approximation a gaussian density matrix in terms
of the Hartree-Fock states. As a consequence of our assumption of no
cross correlation between $\chi$ and $\vec{\pi}$ , the density matrix
factorizes as
\be
\rho(t) = \rho_{\chi}(t)\otimes \rho_{\pi}(t) \nonumber
\ee

In the  Schr\"{o}edinger
representation the density matrix is most easily written down by
making use of spatial translational invariance to
decompose the fluctuation fields $\chi(\x,t)$ and $\vp(\x,t)$ into spatial
Fourier modes:

\ba
\chi(\x,t) & = & \frac{1}{\sqrt{\Omega}} \sum_{\k} \chi_{\k}(t)
\exp(-i\k \cdot \x) \\
\vp(\x)    & = & \frac{1}{\sqrt{\Omega}} \sum_{\k} \vp_{\k}
\exp(-i\k \cdot \x) ,
\ea
where we recall that $\chi(\x,t)=\sigma(\x)-\phi(t)$.
We can now use these Fourier modes as the basis in which to write the density
matrices for the sigma and the pions. We will use the following Gaussian
ansatze:

\ba
\rho_{\chi}[\chi,\tilde{\chi},t] & = & \prod_{\vec{k}}
{\cal{N}}_{\chi,k}(t)
\exp\left\{- \left[\frac{A_{\chi,k}(t)}{2\hbar}\chi_k(t)\chi_{-k}(t)+
\frac{A^*_{\chi,k}(t)}{2\hbar}\tilde{\chi}_k(t)\tilde{\chi}_{-k}(t)+
\right. \right.
\nonumber \\
                                 &   & \left. \left.
 \frac{B_{\chi,k}(t)}{\hbar}\chi_k(t)\tilde{\chi}_{-k}(t)
\right]
+\frac{i}{\hbar}p_{\chi,_k}(t)\left(\chi_{-k}(t)-
\tilde{\chi}_{-k}(t)\right) \right\}, \label{chidensitymatrix}
\ea

\ba
\rho_{\vp}[\vp,\tilde{\vp},t] & = & \prod_{\vec{k}} {\cal{N}}_{\pi,k}(t)
\exp\left\{- \left[\frac{A_{\pi,k}(t)}{2\hbar}\vp_k\cdot \vp_{-k}+
\frac{A^*_{\pi,k}(t)}{2\hbar}\tilde{\vp}_k\cdot \tilde{\vp}_{-k}+
\right. \right.
\nonumber \\
                              &   & \left. \left.
\frac{B_{\pi,k}(t)}{\hbar}\vp_k\cdot
\tilde{\vp}_{-k}\right]\right\} . \label{pidensitymatrix}
\ea
The assumption of isospin invariance implies that the kernels
$A_{\pi,k},\ B_{\pi,k}$
transform as isospin singlets, since these kernels give the two point
 correlation functions. Furthermore, hermiticity of the density matrix
requires that the mixing kernel $B$ be real. The lack of a linear
term in the pion density matrix will become clear below.

The Liouville equation is most conveniently solved in the Schr\"{o}edinger
representation, in which

\[\Pi_{\chi}(\x)= -i\hbar\frac{\delta}
{\delta \chi} \; \; ; \; \; \Pi^j_{\pi}(\x)= -i\hbar\frac{\delta}
{\delta \pi_j} , \]
\be
i \hbar \frac{\partial \rho(t)}{\partial t} =
\left(H[\Pi_{\chi},\vec{\Pi}_{\vp}, \chi,\vp;t]- H[\tilde{\Pi}_{\chi},
\tilde{\vec{\Pi}}_{\vp}, \tilde{\chi},\tilde{\vp};t]\right)\rho(t) .
\ee

Comparing the terms quadratic, linear and independent of the
fields ($\chi \; ; \; \vp$),  we obtain the following set of differential
equations for the coefficients and the expectation value:

\ba
i\frac{{\dot{\cal{N}}}_{\chi,k}}{{\cal{N}}_{\chi,k}} & = &
\frac{1}{2}(A_{\chi,k}-A^*_{\chi,k}) \label{normchi} \\
i\dot{A}_{\chi,k}                                    & = &
\left[A^2_{\chi,k}-B^2_{\chi,k}-\omega^2_{\chi,k}(t) \right]
 \label{achi} \\
i\dot{B}_{\chi,k}                                    & = &
B_{\chi,k}\left(A_{\chi,k}-A^*_{\chi,k}\right) \label{bchi} \\
\omega^2_{\chi,k}(t)                                 & = &
k^2+{\cal{M}}^2_{\chi}(t) \label{omega2chi}
\ea
\be
\ddot\phi+m^2(t)\phi+4\lambda\phi^3+4\lambda\phi \langle \vec{\pi}^2(\x,t)
\rangle -h =  0 .
\label{fieqnofmotion}
\ee

\ba
i\frac{{\dot{\cal{N}}}_{\pi,k}}{{\cal{N}}_{\pi,k}}  & = &
\frac{1}{2}(A_{\chi,k}-A^*_{\chi,k}) \label{normpi} \\
i\dot{A}_{\pi,k}                                    & = &
\left[A^2_{\pi,k}-B^2_{\pi,k}-\omega^2_{\pi,k}(t) \right]
 \label{api} \\
i\dot{B}_{\pi,k}                                    & = &
B_{\pi,k}\left(A_{\pi,k}-A^*_{\pi,k}\right) \label{bpi} \\
\omega^2_{\pi,k}(t)                                 & = &
k^2+{\cal{M}}^2_{\pi}(t). \label{omega2pi}
\ea

The lack of a linear term in (\ref{pidensitymatrix}) is a consequence
of a lack of a linear term in $\vp$ in the Hartree Hamiltonian, as the
symmetry has been specified to be broken along the sigma direction.

To completely solve for the time evolution, we must specify the initial
conditions. We will {\it assume} that at an initial time ($t=0$) the
system is in {\it local thermodynamic equilibrium} at an initial
temperature  $T$, which we take to be higher than the critical
temperature,$T_c \approx \mbox{ 200 MeV }$,  where we use the
phenomenological couplings and masses to obtain $T_c$.

This assumption thus describes the situation in a
high energy collision in which the central rapidity region is at a
temperature larger than critical,
and thus in the symmetric phase, and such that the
phase transition occurs via the rapid cooling that occurs when the
region in the high temperature phase expands along the beam axis.

The assumption of local thermodynamic equilibrium for the Hartree-Fock
states determines the initial values of the kernels and the expectation
value of the sigma field and its canonical momentum:
\ba
A_{\chi,k}(t=0) & = & \omega_{\chi,k}(0)\coth[\beta \hbar
 \omega_{\chi,k}(0)] \label{iniachi} \\
B_{\chi,k}(t=0) & = & -\frac{\omega_{\chi,k}(0)}{\sinh[\beta \hbar
 \omega_{\chi,k}(0)]} \label{inibchi} \\
A_{\pi,k}(t=0)  & = & \omega_{\pi,k}(0)\coth[\beta \hbar
 \omega_{\pi,k}(0)] \label{iniapi} \\
B_{\pi,k}(t=0)  & = & -\frac{\omega_{\pi,k}(0)}{\sinh[\beta \hbar
 \omega_{\pi,k}(0)]} \label{inibpi} \\
\phi(t=0)       & = & \phi_0 \; \; ; \; \;
\dot{\phi}(t=0)  =  0 , \label{initialfidot}
\ea
with $\beta = 1/k_B T$. We have (arbitrarily) assumed that the expectation
value of the canonical momentum conjugate to the sigma field is zero in
the initial equilibrium ensemble. These initial conditions dictate the
following ansatze for the real and imaginary parts
of the kernels $A_{\pi,k}(t),\ A_{\chi,k}(t)$ in terms of
complex functions ${\cal{A}}_{\pi,k}(t)={\cal{A}}_{R;\pi,k}(t)+
i{\cal{A}}_{I;\pi,k}(t)$ and  ${\cal{A}}_{\chi,k}(t)={\cal{A}}_{R;\chi,k}(t)+
i{\cal{A}}_{I;\chi,k}(t)$\cite{boyfrw}:

\ba
A_{R;\pi,k}(t) & = & {\cal{A}}_{R;\pi,k}(t) \coth[\beta \hbar
 \omega_{\pi,k}(0)]  \label{aresol} \\
B_{\pi,k}(t)   & = & -\frac{{\cal{A}}_{R;\pi,k}(t)}{\sinh[\beta \hbar
 \omega_{\pi,k}(0)]}  \label{bsol} \\
A_{I;\pi,k}(t) & = &  {\cal{A}}_{I;\pi,k}(t) \label{aimsol}
\ea
The differential equation for the complex
function ${\cal{A}}$
can be cast in a more familiar form by a change of variables

\be
{\cal{A}}_{\pi,k}(t) = -i \frac{\dot{\Psi}_{\pi,k}(t)}{\Psi_{\pi,k}(t)} ,
\label{psi}
\ee
with $\Psi_{\pi,k}$ obeying the following Schr\"{o}dinger-like differential
equation, and boundary conditions
\ba
\left[\frac{d^2}{dt^2}+\omega^2_{\pi,k}(t) \right]\Psi_{\pi,k}(t)
                         & = & 0 \label{diffeqnpsi} \\
\Psi_{\pi,k}(t=0)        & = & \frac{1}{\sqrt{\omega_{\pi,k}(0)}}
\; \; ; \; \;
\dot{\Psi}_{\pi,k}(t=0)
                          =  i\sqrt{\omega_{\pi,k}(0)} .
\label{psidot0}
\ea

Since in this approximation the dynamics for the pions and sigma fields
decouple, we will only concentrate on the solution for the pion fields;
the effective time dependent frequencies for the sigma fields are
completely determined by the evolution of the pion correlation functions.
In terms of these functions
we finally find
\be
\langle {\vp}_{k}(t) \cdot {\vp}_{-k}(t) \rangle  =
\frac{N\hbar}{2} \mid \Psi_{\pi,k}(t)\mid^2 \coth\left[
\frac{\hbar \omega_{\pi,k}(0)}{2k_BT} \right].
\label{pioncorrfunc}
\ee
In terms of this two-point correlation function and recognizing that
the $\Psi_{\pi,k}(t)$ only depends on $k^2$, we obtain the following
important correlations :

\ba
\langle \vp^2(\x,t) \rangle       & = & \frac{N\hbar}{4\pi^2}\int dk
k^2 \mid \Psi_{\pi,k}(t)\mid^2 \coth\left[
\frac{\hbar \omega_{\pi,k}(0)}{2k_BT} \right] \label{pi2corrfunc} \\
\langle \vp(\x,t) \cdot \vp(\vec{0},t) \rangle
                                  & = & \frac{N\hbar}{4\pi^2}\int dk
k\frac{\sin(kx)}{x} \mid \Psi_{\pi,k}(t)\mid^2 \coth\left[
\frac{\hbar \omega_{\pi,k}(0)}{2k_BT} \right] .
\label{pispatialcorrfunc} \\
\ea
The presence of the temperature dependent function in the above
expressions encodes the finite temperature correlations of the initial
state. The set of equations (\ref{fieqnofmotion},\ref{diffeqnpsi}) with the
above boundary conditions, completely determine the non-equilibrium
 dynamics in the Hartree-Fock approximation. We will provide a numerical
analysis of these equations in the next section.

\subsection{\bf Pion production}

In the Hartree approximation, the Hamiltonian is quadratic, and the
fields can be expanded in terms of creation and annihilation of
Hartree-Fock states
\be
{\vp}_{k}(t)  =  \frac{1}{\sqrt{2}} \left(\vec{a}_k
 \Psi^{\dagger}_{\pi,k}(t)+
\vec{a}^{\dagger}_{-k} \Psi_{\pi,k}(t) \right) .
\label{heisifield}
\ee
The creation $\vec{a}^{\dagger}_k$ and annihilation $\vec{a}_k$
operators are {\it independent of time} in the Heisenberg picture and
the mode functions $\Psi_{\pi,k}(t)$ are the solutions to the Hartree
equations (\ref{diffeqnpsi}) which are the Heisenberg equations of motion
in this approximation. The boundary conditions (\ref{psidot0})
correspond to positive frequency particles for $t \leq 0$.
The creation and annihilation operators may be
written in terms of the Heisenberg fields (\ref{heisifield}) and their
canonical momenta. Passing on to the Schr\"{o}edinger
picture at time $t=0$, we can relate the Schr\"{o}edinger
picture operators at time $t$ to those at time $t=0$ via a
Bogoliubov transformation:
\be
\vec{a}_k (t)  =   {\cal{F}}_{+,k}(t) \vec{a}_k(0) +
{\cal{F}}_{-,k}(t)
\vec{a}^{\dagger}_{-k} (0) \label{bogolitrans}
\ee
with

\ba
\mid {\cal{F}}_{+,k}(t) \mid^2    & = &
\frac{1}{4}\frac{\mid \Psi_{\pi,k}(t) \mid^2}
{\mid \Psi_{\pi,k}(0) \mid^2}
 \left[1+
\frac{\mid {\dot{\Psi}}_{\pi,k}(t)\mid^2}{\omega^2_{\pi,k}(0)
\mid \Psi_{\pi,k}(t) \mid^2}\right] + \frac{1}{2}
 \label{bogolo} \\
\mid {\cal{F}}_{+,k}(t) \mid^2    & - &
\mid {\cal{F}}_{-,k}(t) \mid^2 =1 .
\nonumber
\ea

At any time $t$ the expectation value of the number operator for pions
(in each
$k$-mode) is
\be
\langle N_{\pi,k}(t) \rangle = \frac{Tr \left[\vec{a}^{\dagger}_{\pi,k}(t)\cdot
 \vec{a}_{\pi,k}(t) \rho(0)\right]}{Tr\rho(0)} =
\frac{Tr \left[ \vec{a}^{\dagger}_{\pi,k}(0)\cdot
 \vec{a}_{\pi,k}(0) \rho(t)\right]}{Tr\rho(t)} .
\label{pionnumber}
\ee
After some straightforward algebra we find
\be
\langle N_{\pi,k} \rangle(t) = (2
\mid{\cal{F}}_{+,k}(t,t_o)\mid^2 -1)\langle N_{\pi,k}\rangle(0)+
\left(\mid{\cal{F}}_{+,k}(t,t_o)\mid^2 -1\right) .
\label{numboft}
\ee
The first term represents the ``induced'' and the second term the
``spontaneous'' particle production. In this approximation, particle
production is a consequence of parametric amplification. The
Hartree-Fock states are examples of squeezed states, and the density matrix
is a ``squeezed'' density matrix. The squeeze parameter (the
ratio of the kernels at a time $t$
to those at time $t=0$)  is time
dependent and determines the time evolution of the states and
density matrix. The relation between squeezed states and pion production
has been advanced by Kogan\cite{kogan} although not in the context of an
initial thermal density matrix.

Thus far we have established the formalism to study the non-equilibrium
evolution during the phase transition.
A question of interpretation must be clarified before proceeding further.
Our description, in terms of a statistical density matrix, describes an
isospin invariant mixed state, and thus does not prefer one isospin
direction over another. A real experiment will furnish one realization
of all the available states mixed in the density matrix. However, if
the pion correlation functions become long ranged (as a statistical
average) it is clear that in a particular realization, at least one
isospin component is becoming correlated over large distances, thus it is
in this statistical sense that our results should be understood.

This concludes our discussion of the formalism we will use to study the
non-equilibrium evolution of the pion system. We now turn to a numerical
analysis of the problem.

\subsection{\bf Numerical Analysis}

The phenomenological set of parameters that define the linear
sigma model as an effective low energy theory are as follows
(we will be somewhat cavalier about the precise value of these
parameters as we are interested in the more
robust features of the pion correlations)

\ba
 M_{\sigma} \approx 600 \mbox{ MeV }     & & \; \; ; \; \;
  f_{\pi} \approx
95 \mbox{ MeV } \; \; ; \; \; \lambda \approx 4.5 \nonumber \\
h \approx (120 \mbox{ MeV })^3             & & \; \; ; \; \;
 T_c \approx 200 \mbox{ MeV } .
\label{parameters}
\ea

The above value of the critical temperature differs somewhat from the
lattice estimates ($T_c \approx 150 \mbox{ MeV }$), and our definition of
$\lambda$ differs by a factor four from that given
elsewhere\cite{wilraj2,pisarski}.

The first thing to notice is that this is a {\it strongly} coupled theory,
and unlike our  previous studies of the dynamics of
phase transitions\cite{boyveg,boydom} we expect the relevant time scales
to be much {\it shorter} than in weakly coupled theories.

We must also notice that the linear sigma model is an effective low-energy {\it
cutoff} theory. There are two physically important factors that limit the value
of the cutoff: i) this effective theory neglects the influence of the nucleons,
and in the Hartree approximation, the vector resonances are missed. These two
features imply a cut-off of the order of about 2 GeV; ii) the second issue is
that of the triviality bound. Assuming that the value of the coupling is
determined at energies of the order of $M_{\sigma}$, its very large value
implies that the cutoff should not be much larger than about 4-5 GeV,
since otherwise
the theory will be dangerously close to the Landau pole. From the technical
standpoint this is a more important issue since in order to write the
renormalized equations of motion we need the  ratio between bare and
renormalized couplings. In the Hartree approximation this is the ``wave
function renormalization constant'' for the composite operator ${\vp}^2$.

Thus we use a cutoff $\Lambda=2 \mbox{ GeV }$. The issue of the cutoff
is an important one since $\langle {\vp}^2 \rangle$ requires renormalization,
and in principle we should write down renormalized equations of motion.
In the limit when the cutoff is taken to infinity the resulting evolution
should be insensitive to the cutoff. However the chosen cutoff is not
very much larger than other scales in the problem and the ``renormalized''
equations will yield solutions that are cutoff sensitive. However, this
sensitivity will manifest itself on distance scales of the order of $0.1$ fm
or smaller, and we are interested in detecting correlations over many
fermis. The {\it size} of the correlated regions and the time scales for
their growth will be determined by
the long wavelength unstable modes\cite{boydom} (see below), and
thus should be fairly insensitive to the momentum scales
near the cutoff. The short distance features of the correlation functions,
such as {\it e.g.} the {\it amplitude} of the fluctuations will, however, be
rather sensitive to the cutoff.

The most severe ultraviolet divergence in the composite operator
${\vp}^2$
is proportional to $\Lambda^2$.
This divergence
is usually handled by a subtraction. We will subtract this term
(including the temperature factors) in a renormalization of the mass
at $t=0$.
Thus

\be
m^2_B(t=0)+4\lambda\langle \vp^2 \rangle (t=0;T) =m^2_R(t=0;T)
\label{renormass}
\ee
where we made explicit the temperature dependence of ${\vp}^2$ and
$m^2_R(t=0;T)$.
Furthermore we will parametrize the time dependent mass term as
\be
m^2_R(t)=\frac{M_{\sigma}^2}{2}\left[\frac{T^2}{T^2_c}
\exp{\left[-2\frac{t}{t_r}\right]}-1\right] \Theta(t)+
\frac{M_{\sigma}^2}{2}\left[\frac{T^2}{T^2_c}
-1\right] \Theta(-t) .  \label{masst}
\ee

This parametrization incorporates the dynamics of the expansion and cooling
processes in the plasma in a phenomenological way. It allows for the system
to cool down with an effective temperature given by:

\be
T_{eff}(t) = T\exp{\left[-\frac{t}{t_r}\right]} \label{tempoft}
\ee
where $T$ is the initial value of the temperature in the central rapidity
region, and $t_{r}$ is a relaxation time. This parametrization also allows us
to study a ``quench'' corresponding to the limiting case $t_r=0$.

It is convenient to introduce the natural scale
 $\mbox{1~ fm }^{-1} \approx 200 \mbox{ MeV } = M_F$ and define the following
dimensionless variables
\ba
\phi(t) & = &  M_F f(t) \; \; ; \; \; \Psi_{\pi,k}(t) =
\frac{\psi_q(\tau)}{\sqrt{M_F}} \nonumber \\
 k      & = & M_F q \; \; ; \; \; t =
 \frac{\tau}{M_F} \; \; ; \; \; t_r =
 \frac{\tau_r}{M_F} \; \; ; \; \; x= \frac{z}{M_F} .
\ea

In these units
\ba
\frac{\Lambda}{M_F} & = & 10 \; \; ; \; \; \frac{M_{\sigma}}{M_F} =3
\; \; ; \; \; H=\frac{h}{M_F^3} \approx 0.22
\label{numbers} \\
\frac{\omega^2_{\pi,k}(0)}{M_F^2}
                    & = & W^2_q =
 q^2+\frac{9}{2}\left[\frac{T^2}{T_c^2}
-1\right]+4 \lambda f^2(0) \label{omegadim}
\ea

Thus we have to solve simultaneously the Hartree set of equations:
\ba
 \frac{d^2 f}{d\tau^2}+\frac{9}{2}f\left[\frac{T^2}{T_c^2}
\exp{\left[-2\frac{\tau}{\tau_r}\right]}
-1\right]+4 \lambda f^3 + 4f \lambda \Sigma(0,\tau)-H
& = & 0 \label{dimfieqn} \\
\left\{ \frac{d^2}{d\tau^2}+q^2+\frac{9}{2}\left[\frac{T^2}{T_c^2}
\exp{\left[-2\frac{\tau}{\tau_r}\right]}
-1\right]+4 \lambda f^2(\tau) + 4 \lambda \Sigma(0,\tau) \right\}
 \psi_q(\tau)
& = & 0 \label{dimpsieqn}
\ea
\ba
\Sigma(z,\tau) & = & \langle \vp(\x,t) \cdot \vp(\vec{0},t) \rangle /
 M_F^2 \nonumber \\
               & = & \frac{3}{4\pi^2} \int^{10}_{0}dq q
\frac{\sin(qz)}{z}  (\mid \psi_q(\tau) \mid^2-
\mid \psi_q(0) \mid^2) \coth\left[\frac{W_q}{10 T(\mbox{ GeV })}\right]
\label{Sigma}
\ea
with the boundary conditions
\ba
f(0) & = & f_0 \; \; ; \; \; \frac{df(0)}{d\tau} = 0 \label{fibound} \\
\psi_q(0)
     & = & \frac{1}{\sqrt{W_q}} \; \; ; \; \; \frac{d \psi_q}{d\tau}(0) =
i\sqrt{W_q} \label{psibound}
\ea
Finally, once we find the Hartree mode functions, we can compute the
total pion density as a function of time:
\be
\frac{N_{\pi}(t)}{\Omega} =
\frac{\overline{N}_{\pi}(\tau)}{(\mbox{ fm })^3}
 = \frac{1}{2\pi^2 \mbox{ fm }^3} \int^{10}_{0} dq~
 q^2 \langle N_{\pi,q}(\tau) \rangle \label{piondensity}
\ee
with $\langle N_{\pi,q}(\tau) \rangle $ given by (\ref{numboft}) in terms
of the dimensionless variables.

The mechanism of domain formation and growth is the  fast time evolution
of the unstable modes\cite{boydom}.

 Let us consider first the case of $H=0$. Then $f=0$
is a fixed point of the evolution equation for $f$ and
corresponds to cooling down from the symmetric (disoriented) phase in the
absence of explicit symmetry breaking perturbations. Let us consider
the simpler situation of a quench ($\tau_r=0$). The  equation for the
Hartree mode functions (\ref{dimpsieqn}) shows that for $q^2 < 9/2$ the
corresponding modes are unstable at early times and grow exponentially.

This growth feeds back on $\Sigma(\tau)$ which begins to grow and tends
to overcome the instability. As the unstable fluctuations grow, only longer
wavelengths remain unstable, until the time when
$4 \lambda \Sigma(\tau) \approx 9/2$, at which point no wavelength is unstable.
The modes will continue to grow however,  because the derivatives will be
fairly large, but since the instabilities
will be overcome beyond this time the modes will have an oscillatory behavior.

We expect then that the fluctuations will grow during the time for which
there are instabilities. This time scale depends on the value of the
coupling; for very small coupling, $\Sigma(\tau)$ will have to grow for
a long time before $4 \lambda \Sigma(\tau) \approx 9/2$ and
the instabilities are shut-off. On the other hand, for strong coupling
this time scale will be rather small, and domains will not have much time
to grow.

It is clear that allowing for a non-zero magnetic field or $f(0) \neq 0$
will help to shut off the instabilities at earlier times, thus making
the domains even smaller.

For typical relaxation times ($\approx 1-2 \rm{\mbox{ fm/c }}$),
 domains will not grow too large either because the fast growth of
the fluctuations will  catch up with the relaxing modes and
shut-off the instabilities (when $T_{eff}(\tau) < T_c$) on short time scales.
Thus we expect that for a non-zero  relaxation time $\tau_r \approx
\rm{\mbox{ fm/c }}$, domains will not grow too  large either because
the fluctuations will shut-off the instabilities
(when $T_{eff}(\tau) < T_c$) on shorter time scales (this argument will
be confirmed numerically shortly).

We conclude from this analysis that the optimal situation for which large DCC
regions can grow corresponds to a quench from the symmetric phase in
the absence of a magnetic field.

Figure (9.a) shows $4 \lambda \Sigma(\tau) \mbox{ vs } \tau$,
figure (9.b) shows $\Sigma(z,\tau=1;3;5) \mbox{ vs } z$ and
and figure (9.c) shows $\overline{N}_{\pi}(\tau) \mbox{ vs } \tau$ for
 the
following values of the parameters: $T_c = 200 \mbox{ MeV } \; \; ; \; \;
T = 250 \mbox{ MeV }\; \; ; \; \; \tau_r=0 \; \; ; \; \; f(0)=0 \; \; ;
\; \; H=0 \; \; ; \; \;  \lambda=4.5 $ corresponding to a quench from the
symmetric phase at a temperature slightly above the critical temperature
 and no  magnetic field.

 We clearly see that the fluctuations grow to
overcome the instability in times $\approx 1 \mbox{ fm }$ and the
domains never get bigger than about $\approx 1.5 \mbox{ fm }$.
Figure (1.c) shows that the number of pions per cubic fermi is about
0.15 at the initial time (equilibrium value) and grows to about
0.2 in times about 1-2 fermis after the quench. This pion density is
thus consistent with having only a few pions in a pion-size correlation
volume.

The reason the fluctuations grow so quickly and thus shut off the growth of the
unstable modes so quickly is the strongly coupled nature of the theory.

The possibility of long range correlations exists if the initial
state is in {\it equilibrium at the critical temperature}. In this
situation there are already  long range correlations in the initial
state that will remain for some time as the temperature factors enhance
the contributions for long wavelength modes since the Boltzmann factor
$\approx 1/k$ for long wavelength fluctuations. Figures (10.a-c) show
this situation for the values of the parameters  $ T=T_c = 200
\mbox{ MeV } \; \; ; \; \;
 \tau_r=0 \; \; ; \; \; f(0)=0 \; \; ;
\; \; H=0 \; \; ; \; \;  \lambda=4.5 $. Figure (10.b) shows
 $\Sigma(z,\tau=1;2) \mbox{ vs } z$. In this case the number of
pions per cubic fermi in the initial state is $\approx 0.12$ and
reaches a maximum of about $0.17$ within times of the order of a fermi/c.
 The pions, however, are correlated
over distances of about $4-5 \mbox { fm }$ with a large number of
pions per correlation volume $\approx 50$. These large correlation
volumes are a consequence of the initial long range correlations.
This is the situation proposed by Gavin, Gocksch and Pisarski for the
possibility of formation of large domains, as there is a ``massless''
particle in the initial state.

We believe that this situation is not very likely as the central
rapidity region must remain in equilibrium at (or very close) to the
critical temperature before the quench occurs.

To contrast this situation with that of a weakly coupled theory, figures
(11.a-c) show the same functions with the following values of the parameters
 $ \lambda=10^{-6} \; \; ; \; \; T_c = 200 \mbox{ MeV } \; \; ; \; \;
T = 250 \mbox{ MeV }\; \; ; \; \; \tau_r=0 \; \; ; \; \; f(0)=0 \; \; ;
\; \; H=0$. Now the fluctuations are negligible up to times of about 4
fm/c, during which time the correlation functions grow (figure 11.b)
and pions become correlated over distances of the order 4-5 Fermis. As
shown in figure (11.c) the number of pions per cubic fermi becomes enormous,
a consequence of a large parametric amplification. In this extremely
weakly coupled theory, the situation of a quench from above the critical
temperature to almost zero temperature does produce a large number of
coherent pions and domains which are much larger than typical pion sizes.
This is precisely the situation studied previously\cite{boydom}
 within a different context.

We have analyzed numerically many different situations in the strongly
coupled case ($\lambda \approx 4-5$) including the magnetic field and
letting the expectation value of the sigma field ``roll-down'' etc, and
in {\it all} of these cases in which the initial temperature is higher
than the critical (between $ 10-20 \%$ higher) we find the
 common feature that the time
 and spatial scales of correlations are  $\approx 1 \mbox{ fm }$.  Thus
it seems that within this approach the strongly coupled linear sigma
model is incapable of generating large domains of correlated pions.

\subsection{Discussions and Conclusions:}

Our study differs in many qualitative and quantitave ways from previous
studies. In particular we incorporate both quantum and thermal
fluctuations and correlations in the initial state. In previous studies
it was argued that because one is interested in long-wavelength
fluctuations these may be taken as {\it classical} and the classical
evolution equations (with correlations functions replaced by spatial
averages) were studied. We think that it is important to quantify why
and when the long-wavelength fluctuations are classical within the
present approximation scheme.

This may be seen from the temperature factors in the Hartree propagators. These
are typically (incorporating now the appropriate powers of $\hbar$):
\[\hbar \coth\left[\frac{\hbar \omega_k}{2k_BT}\right] . \]
Thus, modes with wavelength $k$ and energies $\omega_k$ are classical
when
$\hbar \omega_k \ll k_BT$  and yield a contribution to the propagator

\[\hbar \coth\left[\frac{\hbar \omega_k}{2k_BT}\right]
 \approx 2k_B T/ \omega_k\]

(notice the cancellation of the $\hbar$). For long-wavelength components
this happens when

\[ \frac{M_{\sigma}^2}{T^2}[T^2/T^2_c-1] \ll 1 , \]
because the ``thermal mass'' (squared) for the excitations in the
 heat bath is
$\frac{M_{\sigma}^2}{2}[T^2/T^2_c-1]$.
  For the phenomenological values of
$M_{\sigma}$ and $T_c$, the ``classical'' limit is obtained when

\be
[\frac{T^2}{T_c^2}-1] \ll 0.1,
\ee
that is, when the initial state is in {\it equilibrium} at a temperature
that is {\em extremely}
close to the critical temperature. This is the situation that is
shown in figures (3.a-c) where, indeed, we obtain very large correlated
domains that were already present in the initial state after a quench
from the critical temperature all the way to zero temperature.

After an energetic collision it seems rather unlikely that the central
region will be so close to the critical temperature.
If the temperature is higher than critical, in order for the
system to cool down to the critical temperature (or very near to it)
and to remain in {\it local thermodynamic equilibrium},
very long relaxation times are
needed, as the long-wavelength modes are typically critically slowed
down during a transition. Long relaxation times will allow the
fluctuations to shut off the instabilities as they begin to grow and the
system will lose its long range correlations.

This was the original argument that discarded an equilibrium situation
as a candidate for large domains. Furthermore,
 typical heavy ion collisions or high energy processes will
not allow long relaxation times (typically of a few fermis/c). Thus
we believe that in most generic situations, a classical approximation
for the long wavelength modes is not reliable {\it in the Hartree}
approximation.

We should make here a very important point. We are {\it not} saying that
large coherent fluctuations cannot be treated semiclassically. They can.
What we are asserting with the above analysis is that
within the Hartree approximation, long wavelength excitations cannot
be treated as classical. The Hartree approximation in the form used by
these (and most other) authors {\it does not} capture correctly the
physics of coherent semiclassical non-perturbative configurations.

Thus although the most promising situation, within the model under
investigation, is a quench from the critical temperature (or very close
to it) down to zero temperature, it is our impression that this scenario
is physically highly unlikely.

There is another very tantalizing possibility for the formation of
large correlated pion domains within the {\it linear} sigma model and
that is via the creation of a critical droplet that will complete the
phase transition (first order in this case) via the process of thermal
activation over a ``free energy'' barrier.
The small magnetic field (resulting from the small up and down quark
masses) introduces a small metastability\cite{note}. The classical equations of
motion allow for a solution in which the pion field is zero everywhere
and a droplet in the sigma field (this is the O(3) symmetric bounce
responsible of thermal activation in scalar metastable theories in
three dimensions\cite{linde}). Using a spherically symmetric
ansatz for a sigma droplet of radius R
\be
\sigma_{cl}(r) \approx f_{\pi} \tanh[M_{\sigma}(r-R)],
\ee
and assuming, for the sake of argument, that the thin-wall approximation
is reliable, we obtain an approximate form for the energy of the droplet
\be
E \approx 4\pi R^2 f^2_{\pi}M_{\sigma}-\frac{4\pi}{3}R^3
h f_{\pi}.
\ee
The critical radius is thus (this approximation is clearly reliable only as
an  order-of-magnitude estimate) \[ R_c \approx  3-5 \rm{\mbox{ fm }}. \]
Typical free energy barriers are thus of the order of 500-600 MeV.
By considering the fluctuations of the pions around this configuration,
it is conceivable (although we cannot provide a more convincing argument
at this stage) that the unstable mode of the droplet (dilation) that makes
the droplet grow to complete the phase transition via thermal activation,
produces a large amount of correlated pions. This scenario, however,
requires supercooling (the false vacuum to be trapped) which again
requires long relaxation times (again unlikely for strong coupling).

As argued above, this possibility cannot be studied via a Hartree
approximation which
only provides a (select) resummation of  the perturbative expansion and
is probably reliable only for short times, before non-perturbative
configurations truly develop.

Thus we conclude that although our analysis provides a negative answer
to the question of the possibility of large correlated domains near the
chiral phase transition, these results are
valid only within the Hartree approximation
of the linear sigma model. There are several conceivable possibilities
that would have to be studied thoroughly
before any conclusions are reached: i) perhaps the linear sigma model
is not a good candidate for studying the {\it dynamics} of the chiral
phase transition (although it describes the universality class for the
static properties) ii) there are large coherent field configurations
(droplets) that are not captured in the Hartree approximation. This
possibility is rather likely and is closer to the scenario envisaged by
Bjorken, Kowalski and Taylor\cite{dcc5}. These semiclassical coherent
configurations may be responsible for large
regions of correlated pions. An important ingredient in this latter
case must be a deeper understanding of the dynamical (relaxation)
time scales, for which a deeper understanding of the underlying strongly
interacting theory is needed. A particularly relevant question is whether
such a strongly coupled theory can yield enough supercooling so as to
produce such a configuration.

We believe that these two possibilities must be studied further to
give an unequivocal answer to the question of large correlated domains.
We are currently studying the second possibility in more detail.

{\bf Acknowledgements}: D. B. would like to thank D. S. Lee for illuminating
conversations and remarks and the organizers of the Journee'
Cosmologie for a wonderful workshop.
He acknowledges partial support from N.S.F. under
grant award No: PHY-9302534, and a binational collaboration
N.S.F-C.N.R.S. supported
by N.S.F. grant No: INT-9216755.
H. J. de V. is partially supported by the CNRS/NSF binational
program. R. H. was partially supported by D.O.E. under contract
DE-FG02-91-ER40682.

\newpage

{\bf Figure Captions}:

\vspace{1mm}

\bigskip

\underline{Figure 1:} Final contour of evolution,
eventually $T'
\rightarrow \infty$ , $T \rightarrow -\infty$.

\bigskip

\bigskip

\underline{Figure 2(a):} $\eta$ vs $\tau$ (notation in the
text) for
$g= 10^{-7}$, $\eta(0) = 2.3 \times 10^{-5}$, $\eta'(0) =
0$, $L=1$.
The solid line is the classical evolution, the dashed line
is the evolution from the one-loop corrected equation (\ref{eqofmotunst}).

\bigskip

\bigskip

\underline{Figure 2(b):} One-loop contribution including the
coupling $g$ for
the values of the parameters used in Figure 2(a).

\bigskip

\bigskip

\underline{Figure 2(c):} Velocity $\frac{\partial
\eta}{\partial \tau}$, same
values and conventions as in figure 2(a).

\bigskip

\bigskip

\underline{Figure 2(d):} $(q_{max}(\tau))^2$ vs $\tau$ for
the same values as in figure 2(a).

\bigskip

\bigskip

\underline{Figure 3(a):} $\eta$ vs $\tau$ with $g=10^{-7}$
$\eta(0) = 2.27 \times 10^{-5}$, $\eta'(0) = 0$, $L=1$. Same
conventions for solid and dashed lines as in figure 2(a).

\bigskip

\bigskip

\underline{Figure 3(b):} One-loop contribution
to the equations of motion including the
coupling $g$ for the values of the parameters used in Figure 3(a).

\bigskip

\bigskip

\underline{Figure 3(c):} Velocity $\frac{\partial
\eta}{\partial \tau}$, same
values and conventions as in figure 3(a).

\bigskip

\bigskip

\underline{Figure 3(d):} $(q_{max}(\tau))^2$ vs $\tau$ for
the same values as in figure 3(a).

\bigskip

\bigskip

\underline{Figure 4(a):} $\eta$ vs $\tau$ with $g=10^{-7}$
$\eta(0) = 2.258 \times 10^{-5}$, $\eta'(0) = 0$, $L=1$.
Same conventions for  solid and dashed lines as in figure 2(a).

\bigskip

\bigskip

\underline{Figure 4(b):} One-loop contribution including the
coupling $g$ for the values of the parameters used in Figure 4(a).

\bigskip

\bigskip

\underline{Figure 4(c):} Velocity $\frac{\partial
\eta}{\partial \tau}$, same values and conventions as in figure 4(a).

\bigskip

\bigskip

\underline{Figure 4(d):} $(q_{max}(\tau))^2$ vs $\tau$ for
the  same values as in  figure 4(a).

\bigskip

\bigskip

\underline{Figure 5:} Hartree (solid line) and zero
order (dashed line)
results for
$\frac{\lambda_R}{2m^2_f}(\langle\Phi^2(\tau)\rangle-
\langle\Phi^2(0)\rangle) = {\cal{D}}(0,\tau)$,
for $\lambda=10^{-12}$, $\frac{T_i}{T_c}=2$.

\bigskip

\bigskip

\underline{Figure 6.a:} Scaled correlation functions for
$\tau=6$, as function of $x$,
${\cal{D}}^{(HF)}(x,\tau)$ (solid line), and
${\cal{D}}^{(0)}(x,\tau)$ (dashed line).
$\lambda=10^{-12}$, $\frac{T_i}{T_c}=2$.

\bigskip

\bigskip

\underline{Figure 6.b:} Scaled correlation functions for
$\tau=8$,
as function of $x$,
${\cal{D}}^{(HF)}(x,\tau)$ (solid line), and
${\cal{D}}^{(0)}(x,\tau)$
(dashed line).
$\lambda=10^{-12}$, $\frac{T_i}{T_c}=2$.

\bigskip

\bigskip

\underline{Figure 6.c:} Scaled correlation functions for
$\tau=10$, as function of $x$,
${\cal{D}}^{(HF)}(x,\tau)$ (solid line), and
${\cal{D}}^{(0)}(x,\tau)$ (dashed line).
$\lambda=10^{-12}$, $\frac{T_i}{T_c}=2$.

\bigskip

\bigskip

\underline{Figure 6.d:} Scaled correlation functions for
$\tau=12$, as function of $x$,
${\cal{D}}^{(HF)}(x,\tau)$ (solid line), and
${\cal{D}}^{(0)}(x,\tau)$ (dashed line).
$\lambda=10^{-12}$, $\frac{T_i}{T_c}=2$.

\bigskip

\bigskip

\underline{Figure 7:} Hartree (solid line) and zero
order (dashed line)
results for $\frac{\lambda_R}{2m^2_f}(\langle\Phi^2(\tau)
\rangle-\langle\Phi^2(\tau)\rangle) = 3{\cal{D}}(0,\tau)$,
for $\lambda=0.01$, $\frac{T_i}{T_c}=2$.

\bigskip

\bigskip

\underline{Figure 8:} Scaled correlation functions for
$\tau=4.15$, as function of $x$,
${\cal{D}}^{(HF)}(x,\tau)$ (solid line), and
${\cal{D}}^{(0)}(x,\tau)$ (dashed line).
$\lambda=0.01$, $\frac{T_i}{T_c}=2$.

\bigskip

\bigskip

\underline{Figure 9.a:}

 $4 \lambda \Sigma(\tau) \mbox{ vs } \tau$.
 $T_c = 200 \mbox{ MeV } \; \; ; \; \;
T = 250 \mbox{ MeV }\; \; ; \; \; \tau_r=0 \; \; ; \; \; f(0)=0 \; \; ;
\; \; H=0 \; \; ; \; \;  \lambda=4.5 $

\bigskip

\bigskip

\underline{Figure 9.b:}
 $\Sigma(z,\tau=1;3;5) \mbox{ vs } z$ for the same values of the
parameters as in figure (9.a) larger values of time correspond to larger
amplitudes at the origin.

\bigskip

\bigskip

\underline{Figure 9.c:}
$\overline{N}_{\pi}(\tau) \mbox{ vs } \tau$ for the same values of the
parameters as in figure (9.a). $\overline{N}_{\pi}(0)=0.15$.

\bigskip

\bigskip

\underline{Figure 10.a:}

 $4 \lambda \Sigma(\tau) \mbox{ vs } \tau$.
 $T= T_c = 200 \mbox{ MeV }\; \; ; \; \; \tau_r=0
 \; \; ; \; \; f(0)=0 \; \; ;
\; \; H=0 \; \; ; \; \;  \lambda=4.5 $

\bigskip

\bigskip

\underline{Figure 10.b:}
 $\Sigma(z,\tau=1;2) \mbox{ vs } z$ for the same values of the
parameters as in figure (10.a) larger values of time correspond to larger
amplitudes at the origin.

\bigskip

\bigskip

\underline{Figure 10.c:}
$\overline{N}_{\pi}(\tau) \mbox{ vs } \tau$ for the same values of the
parameters as in figure (10.a). $\overline{N}_{\pi}(0)=0.12$.

\bigskip

\bigskip

\underline{Figure 11.a:}

 $4 \lambda \Sigma(\tau) \mbox{ vs } \tau$.
 $T_c = 200 \mbox{ MeV } \; \; ; \; \;
T = 250 \mbox{ MeV }\; \; ; \; \; \tau_r=0 \; \; ; \; \; f(0)=0 \; \; ;
\; \; H=0 \; \; ; \; \;  \lambda=10^{-6}$

\bigskip

\bigskip

\underline{Figure 11.b:}
 $\Sigma(z,\tau=4;6) \mbox{ vs } z$ for the same values of the
parameters as in figure (11.a) larger values of time correspond to larger
amplitudes at the origin.

\bigskip

\bigskip

\underline{Figure 11.c:}
$\overline{N}_{\pi}(\tau) \mbox{ vs } \tau$ for the same values of the
parameters as in figure (11.a). $\overline{N}_{\pi}(0)=0.15$.


\begin{thebibliography}{99}
\bibitem{note} Gavin and M\"{u}ller\cite{gavin} have recently reported
on the possibility that droplets of the metastable phase may be relevant.
\end{thebibliography}
\end{document}